\documentclass[english,showpacs,superscriptaddress,nofootinbib,aps,notitlepage]{revtex4-2}
\usepackage[T1]{fontenc}
\usepackage[latin9]{inputenc}
\usepackage{color}
\usepackage{babel}
\usepackage{bm}
\usepackage{amsmath}
\usepackage{amssymb}
\usepackage{stmaryrd}
\usepackage{graphicx}
\usepackage{wasysym}
\usepackage[pdfusetitle,
 bookmarks=true,bookmarksnumbered=false,bookmarksopen=false,
 breaklinks=false,pdfborder={0 0 1},backref=false,colorlinks=true]
 {hyperref}

\makeatletter

\newcommand{\lyxmathsym}[1]{\ifmmode\begingroup\def\b@ld{bold}
  \text{\ifx\math@version\b@ld\bfseries\fi#1}\endgroup\else#1\fi}

\usepackage{babel}

\makeatother

\begin{document}
\title{Anisotropic flows of light-flavor and charmed hadrons in Pb+Pb collisions
at LHC energy}
\author{Yan-ting Feng}
\affiliation{School of Physics and Physical Engineering, Qufu Normal University,
Shandong 273165, China}
\author{Rui-qin Wang}
\affiliation{School of Physics and Physical Engineering, Qufu Normal University,
Shandong 273165, China}
\author{Feng-lan Shao}
\email{shaofl@mail.sdu.edu.cn}

\affiliation{School of Physics and Physical Engineering, Qufu Normal University,
Shandong 273165, China}
\author{Jun Song }
\email{songjun2011@jnxy.edu.cn}

\affiliation{School of Physics and Electronic Engineering, Jining University, Shandong
273155, China}
\begin{abstract}
We apply a constituent quark equal-velocity combination (EVC) model
to study the elliptic flow ($v_{2}$) and triangular flow ($v_{3}$)
of light-flavor and single-charmed hadrons in Pb+Pb collisions at
$\sqrt{s_{NN}}=$ 2.76 and 5.02 TeV. $v_{2,3}$ of hadrons in the
EVC model can be expressed as a linear superposition of the $v_{2,3}$
of quarks at\textcolor{magenta}{{} }the same velocity as that of the
hadrons. We find that available experimental data for $v_{2}$ and
$v_{3}$ of $p$, $\Lambda$, $\Xi,$ and $\Omega$ as the function
of transverse momentum ($p_{T}$) can be consistently explained by
the EVC formula using a $v_{2}$ of up/down quarks and a $v_{2}$
of strange quarks. In comparison with $v_{2}$ data of $\phi$ at
$\sqrt{s_{NN}}=$2.76\,TeV which can be naturally explained by $v_{2}$
of strange quarks, explanation of data of $\phi$ mesons at $\sqrt{s_{NN}}=$5.02\,TeV
requires an additional contribution of two-kaon coalescence, which
indicates approximately 20\% influence of final-state hadronic rescattering
on $v_{2}$ of $\phi$ at $\sqrt{s_{NN}}=$5.02 TeV. Using $v_{2}$
and $v_{3}$ of light-flavor quarks obtained in studying light-flavor
hadrons and $v_{2}$ and $v_{3}$ of charm quarks determined from
$D^{0}$ meson data, we apply the EVC model to predict the anisotropic
flows of $D_{s}^{+}$, $\Lambda_{c}^{+}$, $\Xi_{c}^{0},$ and $\Omega_{c}^{0}$
and compare them with the available experimental data. The preliminary
data for $v_{2}$ of $D^{0}$, $D_{s}^{+}$, and $\Lambda_{c}^{+}$
at $\sqrt{s_{NN}}=$ 5.36 TeV are found to be naturally described
by the EVC model. 
\end{abstract}
\maketitle

\section{INTRODUCTION\label{sec:Intro} }

In ultra-relativistic heavy-ion collisions, a state of deconfined
quarks and gluons is formed, known as the quark-gluon plasma (QGP)
\citep{Shuryak:1978ij,Shuryak:1980tp}. In non-central collisions,
the initial anisotropic collision geometry causes anisotropic expansion
of the QGP in the transverse momentum plane, which in turn leads to
anisotropic emission of particles in transverse momentum space \citep{Ollitrault:1992bk,Voloshin:1994mz}.
This anisotropic property is measured in experiments by the harmonic
flow coefficients $v_{n}=\left\langle \cos\left[n\left(\varphi-\varPsi_{n}\right)\right]\right\rangle $
in the Fourier expansion of the transverse momentum distribution of
hadrons \citep{Voloshin:1994mz,Poskanzer:1998yz}. 

There have already been many measurements of the anisotropic flow
of particles produced in relativistic heavy-ion collisions at the
Relativistic Heavy Ion Collider (RHIC) and Large Hadron Collider (LHC).
These measurements provide rich and precise data for $v_{2}$, $v_{3}$,
and higher-order flows of light-flavor and heavy-flavor hadrons. These
data and the relevant theoretical studies reveal lots of important
physical properties of the QGP created in ultra-relativistic heavy-ion
collisions \citep{Kolb:2003zi,Kolb:2000sd,Huovinen:2001cy,Teaney:2000cw,STAR:2013ayu}.
An interesting phenomenon related to the hadronization of the QGP
is that $v_{2}$ data of light-flavor hadrons (e.g., $\pi,K,p,\Lambda$)
exhibit a remarkable number-of-constituent-quark (NCQ) scaling \citep{PHENIX:2006dpn,STAR:2015gge,ALICE:2014wao,ALICE:2018yph,STAR:2013ayu}.
This scaling behavior is an indication of the quark recombination
mechanism at QGP hadronization. The NCQ effect is particularly pronounced
in the intermediate transverse momentum region ($2\apprle p_{T}\lesssim6$
GeV/c), where the enhanced baryon-to-meson ratio is also observed
and can be naturally explained by the quark combination models \citep{Greco:2003xt,Chang:2023zbe,Fries:2003vb,Greco:2003mm,Fries:2003kq,Molnar:2003ff,Minissale:2015zwa}.
Recent measurements of charmed hadrons (such as $D^{0}$, $D_{s}^{+}$,
and $\Lambda_{c}^{+}$) in Pb+Pb collisions by CMS and ALICE collaborations
have found that their anisotropic flow exhibits similar behavior to
that of light-flavor hadrons \citep{CMS:2020bnz,ALICE:2020pvw,ALICE:2021kfc,CMS:2025cdf,Bailhache:2014fia,Saha:2024mmw,Saha:2025hsl},
suggesting that charm quarks may undergo partial thermalization and
participate in the collective expansion of the QGP. 

The ALICE and CMS collaborations have reported the precise measurement
of the $v_{2}$ and $v_{3}$ of light-flavor and single-charmed hadrons
in Pb+Pb Collisions at the LHC energies \citep{ALICE:2014wao,ALICE:2022zks,ALICE:2018yph,Zhu:2019twz,CMS:2025cdf,CMS:2020bnz,ALICE:2020pvw,ALICE:2021kfc}.
These experimental data have been studied in various theoretical models
and event generators such as AMPT model \citep{Mallick:2023vgi,Tang:2025fma},
the viscous hydrodynamic model \citep{Alqahtani:2017tnq,Alalawi:2021jwn},
the VISHNU hybrid model \citep{Zhu:2016qiv}, the Boltzmann transport
equation \citep{Akhil:2023xpb}, the blast\nobreakdash-wave\nobreakdash-Tsallis\nobreakdash-power
model (BWTPM) \citep{Grigoryan:2025kba}, and the HYDJET++ model \citep{Devi:2023wih,Devi:2024cxy}.
In this paper, we apply an equal-velocity quark combination model
to systematically study $v_{2}$ and $v_{3}$ of light-flavor and
single-charmed hadrons in Pb+Pb Collisions at $\sqrt{s_{NN}}=$ 2.76
and 5.02 TeV and put particular emphasis on the self-consistency of
the quark combination mechanism in explaining the experimental data
of $v_{2}$ and $v_{3}$ of different hadrons measured by the CMS
and ALICE collaborations. The equal-velocity combination of constituent
quarks gives simple formulas for hadronic flow by the linear superposition
of flow of quarks, which are convenient to be tested by experimental
data and meanwhile are also quite visible in explaining/exhibiting
the correlations among flow of different hadrons. 

The paper is organized as follows. In Sec.~\ref{sec:EVC_model},
we present the analytical formulas of the anisotropic flow hadrons
in a constituent quark equal-velocity combination model and show the
$v_{2}$ and $v_{3}$ of hadrons as the linear superposition of quark
flows. In Sec.~\ref{sec:light hadron}, we apply the EVC model to
study the $v_{2}$ and $v_{3}$ of light-flavor hadrons $\phi$, $\Lambda$,
$\Xi$, $\Omega$, and $p$ in different centralities at $\sqrt{s_{NN}}=$
2.76 and 5.02 TeV. In Sec.~\ref{sec:charm hadron}, we study the
$v_{2}$ and $v_{3}$ of single-charmed hadrons $D^{0}$, $D_{s}^{+}$,
$\Lambda_{c}^{+}$, $\Xi_{c}^{0}$, and $\Omega_{c}^{0}$, and discuss
the scaling property of $v_{3}$ of light-flavor quarks and charm
quarks. Finally, the summary is given in Sec.~\ref{sec:Summary}. 

\section{the anisotropic flows of hadrons in QCM with EVC\label{sec:EVC_model}}

In this section, we employ a constituent quark combination model (QCM)
under the equal-velocity combination (EVC) approximation\citep{Song:2020uvu,Chang:2023zbe,Song:2017gcz,Feng:2025wde,Wang:2024hok}
to describe the production of single-charmed hadrons and derive their
elliptic flow ($v_{2}$) and triangular flow ($v_{3}$) at mid-rapidity
in heavy-ion collisions. In the EVC approximation, the hadron is formed
by the combination of constituent quarks and antiquarks with the same
velocity as that of the hadron. In this ideal case, the momentum vectors
of quarks are parallel to those of hadrons they form, and we have
$\bm{p}_{q}=x_{q}\bm{p}_{h}$ where $x_{q}$ is momentum fraction
of quark $q$ in the hadron $h$. For the production of hadrons at
mid-rapidity, we define the distribution function $f\left(p_{T},\varphi\right)\equiv dN/dp_{T}d\varphi$
where $p_{T}$ is transverse momentum and $\varphi$ is the azimuthal
angle. EVC mechanism constrains $p_{T,q}=x_{q}p_{T,h}$ and $\varphi_{q}=\varphi_{h}$.
Following the spirit of EVC model, the momentum distribution function
of single-charmed hadrons can be expressed as
\begin{align}
f_{M_{j}}\left(p_{T},\varphi\right) & =\kappa_{M_{j}}f_{q_{1}}\left(x_{1}p_{T},\varphi\right)f_{\bar{q}_{2}}\left(x_{2}p_{T},\varphi\right),\label{eq:fmi_indep}\\
f_{B_{j}}\left(p_{T},\varphi\right) & =\kappa_{B_{j}}f_{q_{1}}\left(x_{1}p_{T},\varphi\right)f_{q_{2}}\left(x_{2}p_{T},\varphi\right)f_{q_{3}}\left(x_{3}p_{T},\varphi\right).\label{eq:fbi_indep}
\end{align}
Here, the coefficients $\kappa_{M_{j}}$ and $\kappa_{B_{j}}$ are
independent of the transverse momentum $p_{T}$. The momentum fractions
are defined as $x_{1,2}=m_{1,2}/\left(m_{1}+m_{2}\right)$ for mesons
$M_{j}\left(q_{1}\bar{q}_{2}\right)$, under the momentum conservation
condition $x_{1}+x_{2}=1,$ and $x_{1,2,3}=m_{1,2,3}/\left(m_{1}+m_{2}+m_{3}\right)$
for baryons $B_{j}\left(q_{1}q_{2}q_{3}\right)$, under the momentum
conservation condition $x_{1}+x_{2}+x_{3}=1$. The constituent masses
of quarks are adopted as $m_{u}=m_{d}=0.3$ GeV, $m_{s}=0.5$ GeV,
and $m_{c}=1.5$ GeV, respectively.

The azimuthal dependence of particle distribution function is expressed
through the Fourier series expansion
\begin{equation}
f\left(p_{T},\varphi\right)=f\left(p_{T}\right)\biggl[1+2\sum_{n=1}^{\infty}v_{n}\left(p_{T}\right)\cos(n\varphi)\biggr],\label{eq:fq}
\end{equation}
where $f_{q}\left(p_{T}\right)$ is independent of the azimuthal angle
$\varphi$. The anisotropic coefficient $v_{n}$ is calculated by
\begin{equation}
v_{n}\left(p_{T}\right)=\frac{\int f\left(p_{T},\varphi\right)\cos(n\varphi)d\varphi}{\int f\left(p_{T},\varphi\right)d\varphi}.\label{eq:vn}
\end{equation}

Using Eqs. (\ref{eq:fmi_indep}$-$\ref{eq:fq}), we can obtain the
$v_{n}\left(p_{T}\right)$ for the meson $M_{j}\left(q_{1}\bar{q}_{2}\right)$
and baryon $B_{j}\left(q_{1}q_{2}q_{3}\right)$ and the expressions
up to order $v_{n}^{3}$ are 
\begin{align}
v_{n,M_{j}} & =\frac{\int f_{M_{j}}\left(p_{T},\varphi\right)\cos(n\varphi)d\varphi}{\int f_{M_{j}}\left(p_{T},\varphi\right)d\varphi}\nonumber \\
 & =v_{n,q_{1}}\left[1+\sum_{k=1}^{\infty}\frac{v_{k,q_{1}}}{v_{n,q_{1}}}v_{n+k,\overline{q}_{2}}-2\sum_{k=1}^{\infty}v_{k,q_{1}}v_{k,\overline{q}_{2}}\right]\nonumber \\
 & +v_{n,\overline{q}_{2}}\left[1+\sum_{k=1}^{\infty}\frac{v_{k,\overline{q}_{2}}}{v_{n,\overline{q}_{2}}}v_{n+k,q_{1}}-2\sum_{k=1}^{\infty}v_{k,q_{1}}v_{k,\overline{q}_{2}}\right]\nonumber \\
 & +\sum_{k=1}^{n-1}v_{k,q_{1}}v_{n-k,\overline{q}_{2}}+\mathcal{O}(v^{4}),\label{eq:vn_M_full}
\end{align}
and 
\begin{align}
 & v_{n,B_{j}}\nonumber \\
 & =v_{n,q_{1}}\left\{ 1+\sum_{k=1}^{\infty}\frac{v_{k,q_{1}}}{v_{n,q_{1}}}\left(v_{n+k,q_{2}}+v_{n+k,q_{3}}\right)-2\sum_{k=1}^{\infty}\left(v_{k,q_{2}}v_{k,q_{1}}+v_{k,q_{2}}v_{k,q_{3}}+v_{k,q_{1}}v_{k,q_{3}}\right)\right.\nonumber \\
 & \left.+\sum_{k=1}^{\infty}\sum_{m=1}^{\infty}\frac{v_{k,q_{1}}}{v_{n,q_{1}}}v_{m,q_{2}}\left(v_{m+k+n,q_{3}}+v_{m+k-n,q_{3}}\right)\right\} \nonumber \\
 & +v_{n,q_{2}}\left\{ 1+\sum_{k=1}^{\infty}\frac{v_{k,q_{2}}}{v_{n,q_{2}}}\left(v_{n+k,q_{3}}+v_{n+k,q_{1}}\right)-2\sum_{k=1}^{\infty}\left(v_{k,q_{2}}v_{k,q_{1}}+v_{k,q_{2}}v_{k,q_{3}}+v_{k,q_{1}}v_{k,q_{3}}\right)\right.\nonumber \\
 & \left.+\sum_{k=1}^{\infty}\sum_{m=1}^{\infty}\frac{v_{k,q_{2}}}{v_{n,q_{2}}}v_{m,q_{3}}\left(v_{m+k+n,q_{1}}+v_{m+k-n,q_{1}}\right)\right\} \nonumber \\
 & +v_{n,q_{3}}\left\{ 1+\sum_{k=1}^{\infty}\frac{v_{k,q_{3}}}{v_{n,q_{3}}}\left(v_{n+k,q_{1}}+v_{n+k,q_{2}}\right)-2\sum_{k=1}^{\infty}\left(v_{k,q_{2}}v_{k,q_{1}}+v_{k,q_{2}}v_{k,q_{3}}+v_{k,q_{1}}v_{k,q_{3}}\right)\right.\nonumber \\
 & \left.+\sum_{k=1}^{\infty}\sum_{m=1}^{\infty}\frac{v_{k,q_{3}}}{v_{n,q_{3}}}v_{m,q_{1}}\left(v_{m+k+n,q_{2}}+v_{m+k-n,q_{2}}\right)\right\} \nonumber \\
 & +\sum_{k=1}^{n-1}\left(v_{k,q_{1}}v_{n-k,q_{2}}+v_{k,q_{2}}v_{n-k,q_{3}}+v_{k,q_{3}}v_{n-k,q_{1}}\right)+\sum_{k,m=1}^{m+k\leq n-1}v_{k,q_{1}}v_{m,q_{2}}v_{n-m-k,q_{3}}+\mathcal{O}(v^{4}).\label{eq:vn_B_full}
\end{align}
In the above formulas, we use the abbreviation $v_{n,q_{i}}$ for
$v_{n,q_{i}}\left(x_{i}p_{T}\right)$ with $i=1,2,3$ and $v_{n,\bar{q}_{2}}$
for $v_{n,\bar{q}_{2}}\left(x_{2}p_{T}\right)$.
\begin{widetext}
From Eqs. (\ref{eq:vn_M_full}) and (\ref{eq:vn_B_full}), we see
that $v_{n}$ of hadrons have a general structure of the summation
of $v_{n}$ of quarks and the product of lower-order quark flows.
There is a modification factor before $v_{n,q}$ which contains complex
contributions of flows of quarks at various orders even for low order
flows of hadrons. In order to obtain simple formulas for hadronic
flow at mid-rapidity, we should consider some approximations based
on the properties of hadronic flow measured in relativistic heavy-ion
collisions. 

First, the experimental measurements show a very small value of $v_{1,h}\leq10^{-3}$
at mid-rapidity in heavy-ion collisions at both RHIC and LHC energies,
\citep{STAR:2008jgm,ALICE:2013xri}, which indicates $v_{1,q}\approx(\frac{1}{2}\sim\frac{1}{3})v_{1,h}\leq10^{-3}$
and can be neglected in our analysis. Second, the rough estimation
for $v_{2-5,q}\approx(\frac{1}{2}\sim\frac{1}{3})v_{2-5,h}$ is approximately
$10^{-2}$ according to experimental data of $v_{2-5,h}$ \citep{PHENIX:2006dpn,STAR:2015gge,ALICE:2014wao,ALICE:2018yph,STAR:2013ayu}.
Higher-power terms such as $\left(v_{n,q}\right)^{4,5}$ become negligible
(\ensuremath{\le}$10^{-8}$) and can be safely omitted from the calculations.
Furthermore, we adopt a scaling relation $v_{n,q}=a_{n}v_{2,q}^{n/2}$
at the quark level, which has been indicated by both experimental
and theoretical studies \citep{STAR:2004jwm,STAR:2007afq,Kolb:2004gi,Chen:2004dv,Gardim:2012yp,Borghini:2005kd,Gombeaud:2009ye}.
Using this relation, we can estimate that the ratio $v_{n+1,q_{1}}v_{n+1,q_{2}}/v_{n,q_{1}}v_{n,q_{2}}\sim10^{-2}$
is very small. Consequently, correction terms with harmonic indices
$k,m>2$ in the modification factor of $v_{n,q_{i}}$ can be safely
neglected. The detailed derivation of these approximations is given
in Ref. \citep{Feng:2025wde}. 

Based on the above considerations, the $v_{2}$ and $v_{3}$ of hadrons
can be significantly simplified as follows
\begin{align}
v_{2,M_{j}} & \approx v_{2,q_{1}}\left[1+\left(a_{4}-2\right)v_{2,q_{1}}v_{2,\overline{q}_{2}}\right]+v_{2,\overline{q}_{2}}\left[1+\left(a_{4}-2\right)v_{2,q_{1}}v_{2,\overline{q}_{2}}\right],\label{eq:v2m_final}
\end{align}
\begin{align}
v_{2,B_{j}} & \approx v_{2,q_{1}}\left\{ 1+\left(a_{4}-2\right)\left(v_{2,q_{1}}v_{2,q_{2}}+v_{2,q_{1}}v_{2,q_{3}}\right)-2v_{2,q_{2}}v_{2,q_{3}}\right\} \nonumber \\
 & +v_{2,q_{2}}\left\{ 1+\left(a_{4}-2\right)\left(v_{2,q_{1}}v_{2,q_{2}}+v_{2,q_{2}}v_{2,q_{3}}\right)-2v_{2,q_{1}}v_{2,q_{3}}\right\} \nonumber \\
 & +v_{2,q_{3}}\left\{ 1+\left(a_{4}-2\right)\left(v_{2,q_{1}}v_{2,q_{3}}+v_{2,q_{2}}v_{2,q_{3}}\right)-2v_{2,q_{2}}v_{2,q_{1}}\right\} ,\label{eq:v2b_final}
\end{align}
\begin{align}
v_{3,M_{j}} & \approx v_{3,q_{1}}\left[1+\left(\frac{a_{5}}{a_{3}}-2\right)v_{2,q_{1}}v_{2,\overline{q}_{2}}\right]+v_{3,\overline{q}_{2}}\left[1+\left(\frac{a_{5}}{a_{3}}-2\right)v_{2,q_{1}}v_{2,\overline{q}_{2}}\right],\label{eq:v3m_final}
\end{align}
\begin{align}
v_{3,B_{j}} & \approx v_{3,q_{1}}\left\{ 1+\left(\frac{a_{5}}{a_{3}}-2\right)\left(v_{2,q_{1}}v_{2,q_{2}}+v_{2,q_{1}}v_{2,q_{3}}\right)-2v_{2,q_{2}}v_{2,q_{3}}\right\} \nonumber \\
 & +v_{3,q_{2}}\left\{ 1+\left(\frac{a_{5}}{a_{3}}-2\right)\left(v_{2,q_{1}}v_{2,q_{2}}+v_{2,q_{2}}v_{2,q_{3}}\right)-2v_{2,q_{1}}v_{2,q_{3}}\right\} \nonumber \\
 & +v_{3,q_{3}}\left\{ 1+\left(\frac{a_{5}}{a_{3}}-2\right)\left(v_{2,q_{1}}v_{2,q_{3}}+v_{2,q_{2}}v_{2,q_{3}}\right)-2v_{2,q_{2}}v_{2,q_{1}}\right\} .\label{eq:v3b_final}
\end{align}
\end{widetext}

\section{THE ANISOTROPY FLOW OF LIGHT-FLAVOR HADRONS\label{sec:light hadron} }

In this section, we employ the formulas of anisotropic flow of hadrons
in EVC model presented in above section to study the $p_{T}$ dependence
of the $v_{2}$ and $v_{3}$ for light-flavor hadrons $\phi$, $\Lambda\left(\Lambda+\bar{\Lambda}\right)$,
$\Xi\left(\Xi^{-}+\bar{\Xi}^{+}\right)$, $\Omega\left(\Omega^{-}+\bar{\Omega}^{+}\right)$,
and $p\left(p+\bar{p}\right)$ at mid-rapidity in different centralities
in Pb+Pb collisions at $\sqrt{s_{NN}}=$ 2.76 and 5.02 TeV. 

\subsection{THE ELLIPTIC FLOW OF LIGHT-FLAVOR HADRONS\label{sec:light hadron_v2} }

In the $v_{2}$ of light-flavor hadrons in Eqs. (\ref{eq:v2m_final})
and (\ref{eq:v2b_final}), there is the contribution from the quadratic
term of quark $v_{2}$. In previous study \citep{Feng:2025wde}, we
estimated $a_{4}\approx2$ using the experimental data of $v_{2}$
of hadrons in Au+Au collisions at $\sqrt{s_{NN}}=$200\,GeV \citep{PHENIX:2006dpn,STAR:2015gge,ALICE:2014wao,ALICE:2018yph,STAR:2013ayu}.
The magnitude of $(a_{4}-2)v_{2,q}^{2}$ in the modification term
is therefore very small, and its contribution can be safely neglected.
Consequently, the expression for the $v_{2}$ of light-flavor hadrons
can be simplified as follows
\begin{align}
v_{2,M_{j}}\left(p_{T}\right) & \approx v_{2,q_{1}}\left(x_{1}p_{T}\right)+v_{2,\bar{q}_{2}}\left(x_{2}p_{T}\right),\label{eq:v2,Mi_simp}\\
v_{2,B_{j}}\left(p_{T}\right) & \approx v_{2,q_{1}}\left(x_{1}p_{T}\right)+v_{2,q_{2}}\left(x_{2}p_{T}\right)+v_{2,q_{3}}\left(x_{3}p_{T}\right).\label{eq:v2,Bi_simp}
\end{align}

Using the simplified expressions from Eqs. (\ref{eq:v2,Mi_simp})
and (\ref{eq:v2,Bi_simp}), we obtain the following analytic formulas
for the $v_{2}$ of the various hadrons
\begin{align}
v_{2,p}\left(p_{T}\right) & =3v_{2,u}\left(p_{T}/3\right),\label{eq:v2,p}\\
v_{2,\Omega}\left(p_{T}\right) & =3v_{2,s}\left(p_{T}/3\right),\label{eq:v2,omg}\\
v_{2,\phi}\left(p_{T}\right) & =2v_{2,s}\left(p_{T}/2\right),\label{eq:v2,phi}\\
v_{2,\Lambda}\left(p_{T}\right) & =2v_{2,u}\bigl(\frac{1}{2+r_{su}}p_{T}\bigr)+v_{2,s}\bigl(\frac{r_{su}}{2+r_{su}}p_{T}\bigr),\label{eq:v2,lamb}\\
v_{2,\Xi}\left(p_{T}\right) & =2v_{2,s}\bigl(\frac{r_{su}}{1+2r_{su}}p_{T}\bigr)+v_{2,u}\bigl(\frac{1}{1+2r_{su}}p_{T}\bigr).\label{eq:v2,xi}
\end{align}
Here, we assume the approximate isospin symmetry between up and down
quarks, i.e., $v_{2,u}=v_{2,d}$, and strangeness neutrality, i.e.,
$v_{2,s}=v_{2,\bar{s}}$, in the mid-rapidity region. The relative
momentum ratio $r_{su}=m_{s}/m_{u}$ represents the constituent mass
ratio between strange and up quarks. Considering uncertainties in
constituent quark masses, e.g., $m_{s}$$=0.5\lyxmathsym{\textendash}0.55$
GeV and $m_{u}=0.3\lyxmathsym{\textendash}0.33$ GeV, the derived
ratio $r_{su}$ retains an approximate value of $5/3$, with an estimated
uncertainty of approximately 10\%. Our numerical results show that
the $v_{2}$ of hadrons is not sensitive to such variations in $r_{su}$.
Consequently, this study focuses exclusively on presenting calculated
$v_{2}$ for hadrons using a fixed value of $r_{su}$$=1.67$, which
serves as a representative benchmark for our model predictions. 

\begin{figure}
\centering{}\includegraphics[width=0.65\linewidth,viewport=0bp 00bp 620bp 480bp]{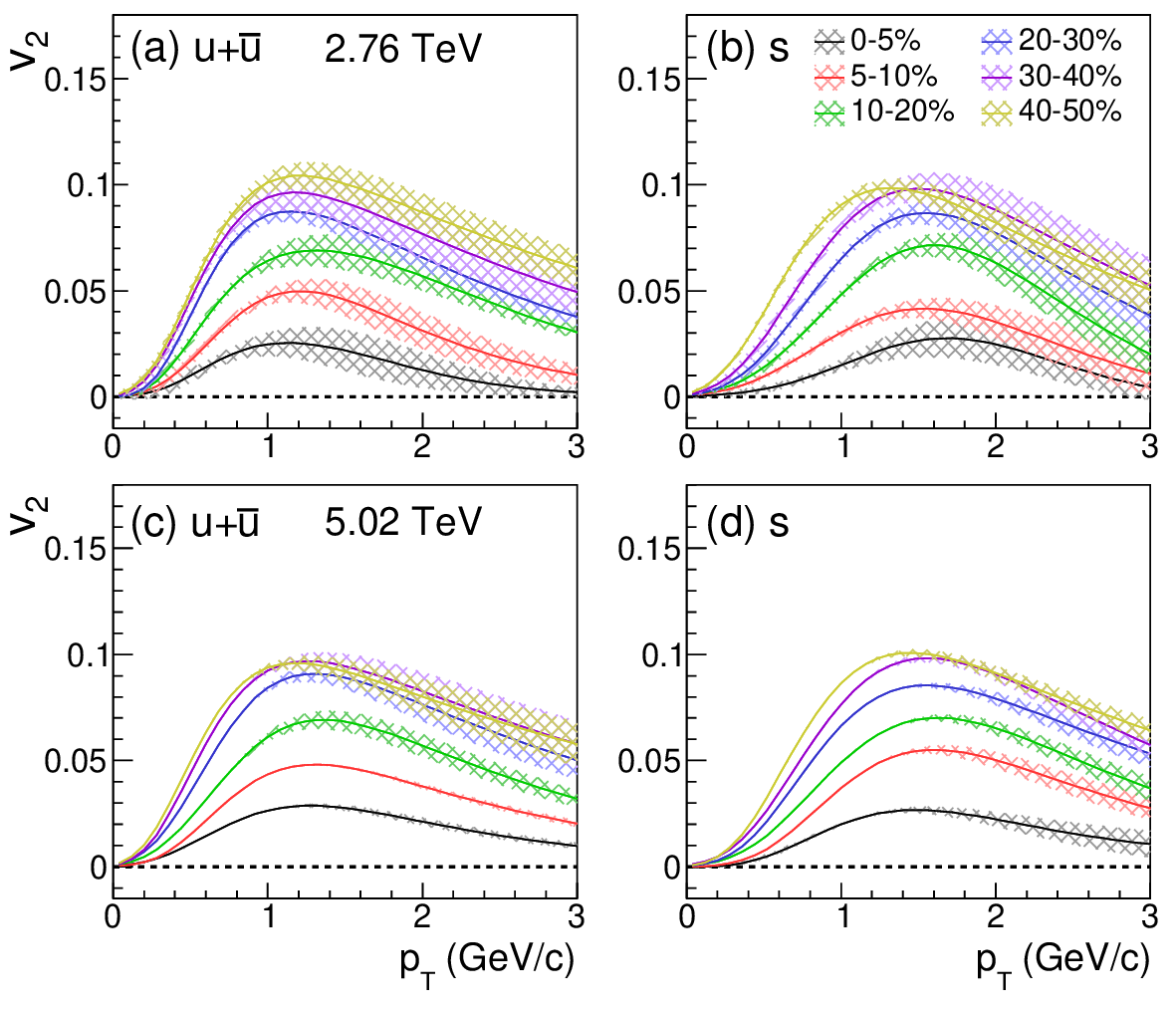}\caption{$v_{2}$ of quarks as a function of $p_{T}$ in different centralities
in Pb+Pb collisions at $\sqrt{s_{NN}}=$2.76 and 5.02 TeV. \label{fig:fig1}}
\end{figure}

In Eqs. (\ref{eq:v2,p})-(\ref{eq:v2,xi}), we have only two inputs
$v_{2,u}$ and $v_{2,s}$, which can be fixed by fitting the experimental
data for $v_{2}$ of $\Lambda$ and $\Xi$ \citep{ALICE:2014wao,ALICE:2022zks}
by Eq. (\ref{eq:v2,lamb}) and (\ref{eq:v2,xi}), respectively. If
data of $\Xi$ are unavailable such as those in the 0--5\% and 5--10\%
centralities at $\sqrt{s_{NN}}=$ 5.02 TeV, the values of $v_{2,u}$
and $v_{2,s}$ are extracted using the available data for $\Lambda$
and $p$ in these centralities. $v_{2}$ of quarks is parameterized
as 
\begin{equation}
v_{2,q}\left(p_{T}\right)=a_{q}\exp\left[-\frac{p_{T}}{b_{q}}-c_{q}\exp\left(-\frac{p_{T}}{d_{q}}\right)\right],\label{eq:v2q}
\end{equation}
where $a_{q},$ $b_{q},$ $c_{q}$, and $d_{q}$ are parameters that
control the shape of $v_{2}$ of quarks. The extracted results for
the $v_{2}$ of quarks in central 0-5\% to semi-central 40-50\% in
Pb+Pb collisions at $\sqrt{s_{NN}}=$2.76 and 5.02 TeV are shown in
Fig.~\ref{fig:fig1}. The shadow areas show the statistical uncertainties,
which are passed from the statistical uncertainties of the experimental
data for $\Lambda$, $\Xi$ or $\Lambda$, $p$.

\begin{figure}
\centering{}\includegraphics[width=0.85\linewidth,viewport=0bp 00bp 580bp 340bp]{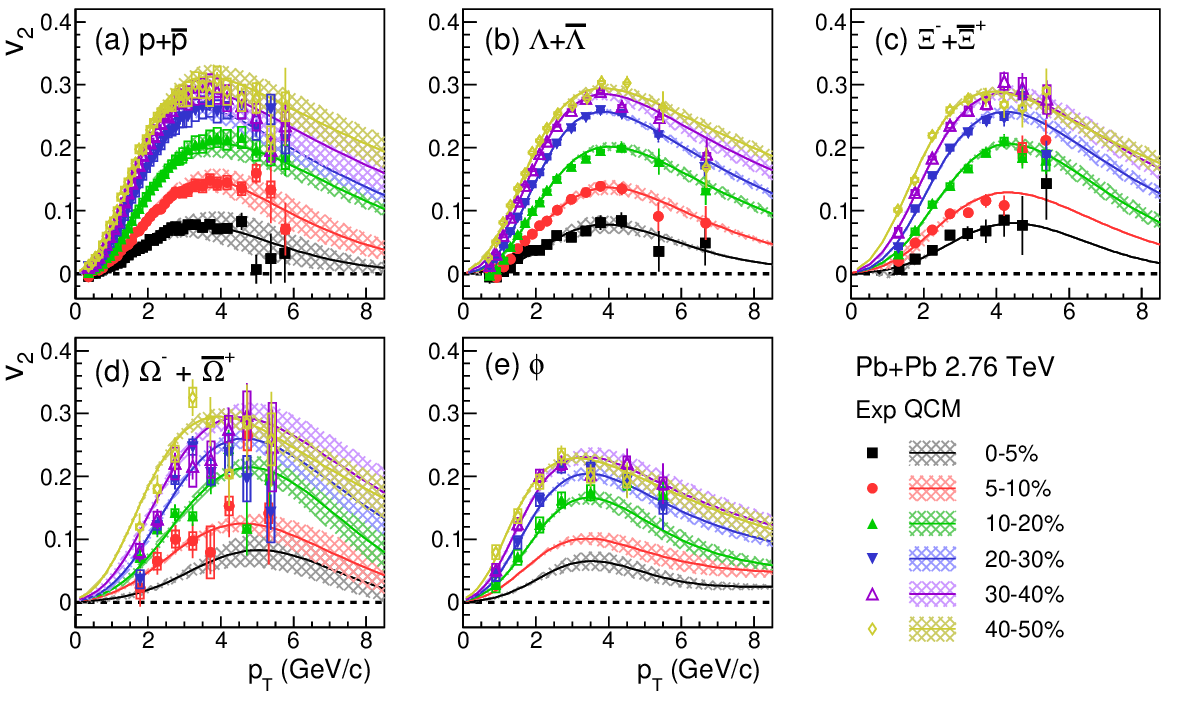}\caption{The $v_{2}$ of (a) $p$, (b) $\Lambda$, (c) $\Xi$, (d) $\Omega,$
and (e) $\phi$ in different centralities in Pb+Pb collisions at $\sqrt{s_{NN}}=$2.76
TeV. Symbols are the experimental data \citep{ALICE:2014wao} and
lines are model results. \label{fig:fig2}}
\end{figure}

\begin{figure}
\centering{}\includegraphics[width=0.85\linewidth,viewport=0bp 00bp 580bp 330bp]{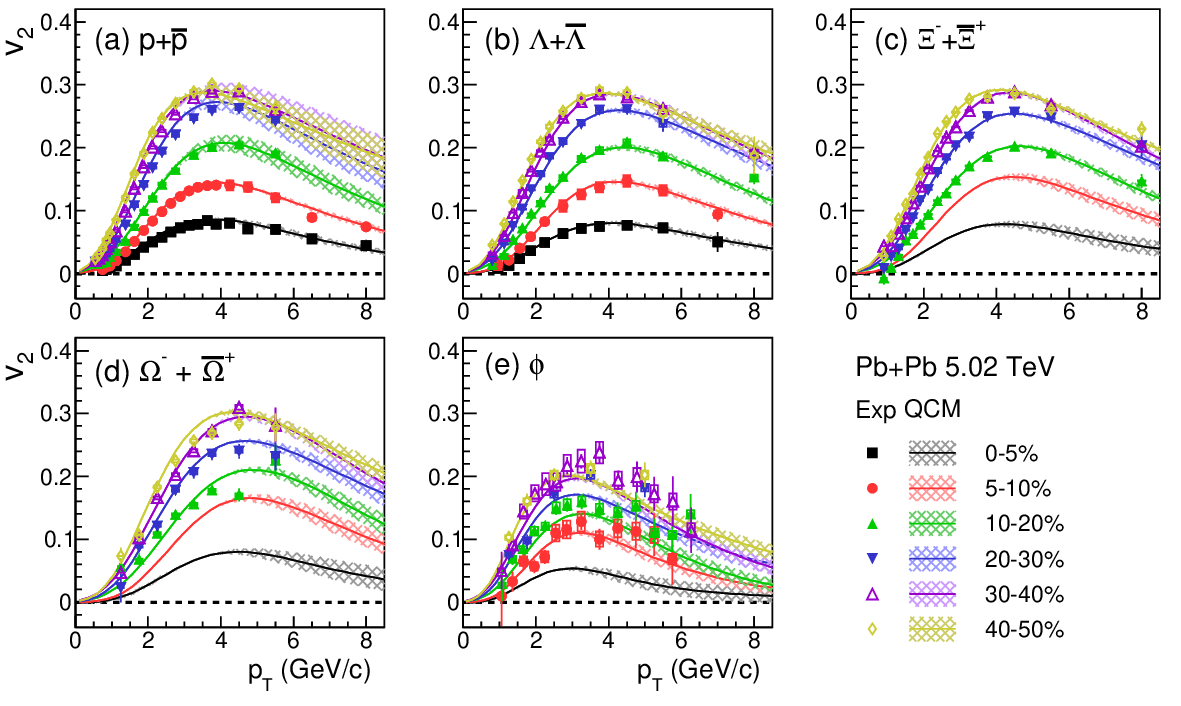}\caption{The $v_{2}$ of (a) $p$, (b) $\Lambda$, (c) $\Xi$, (d) $\Omega,$
and (e) $\phi$ in different centralities in Pb+Pb collisions at $\sqrt{s_{NN}}=$5.02
TeV. Symbols are the experimental data \citep{ALICE:2022zks,Zhu:2019twz}
and lines are model results.\label{fig:fig3}}
\end{figure}

In Figs.~\ref{fig:fig2}$-$\ref{fig:fig3}, we show the results
for the $v_{2}$ of the light-flavor hadrons $\phi$, $\Lambda$,
$\Xi$, $\Omega,$ and $p$ at mid-rapidity in different centralities
(0-5\%, 5-10\%, 10-20\%, 20-30\%, 30-40\%, and 40-50\%) and compare
them with the experimental data in Pb+Pb collisions at $\sqrt{s_{NN}}=$2.76
and 5.02 TeV. The QCM results are lines and experimental data \citep{ALICE:2014wao,ALICE:2022zks}
are symbols. The experimental data of $\Lambda$ and $\Xi$ in the
10-20\%, 20-30\%, 30-40\%, and 40-50\% centralities at $\sqrt{s_{NN}}=$2.76
and 5.02 TeV are used to constrain the quark $v_{2}$, while in 0--5\%
and 5--10\% centralities at $\sqrt{s_{NN}}=$ 5.02 TeV, the data
for $\Lambda$ and $p$ are used. The $v_{2}$ of other hadrons are
theoretical predictions.

We see that the agreement between QCM results for light-flavor hadrons
and experimental data is generally well. An exception is $\phi$ mesons
at $\sqrt{s_{NN}}=$ 5.02 TeV where theoretical results are below
the experimental data to a certain extent. This indicates other production
channels may contribute to $\phi$ mesons production in heavy-ion
collisions at $\sqrt{s_{NN}}=$ 5.02 TeV. Here, we consider the coalescence
of two kaons ($KK\rightarrow\phi$) at the hadronic rescattering stage
\citep{ALICE:2013xri,NA49:2008goy,Sun:2011kj} as a new channel of
$\phi$ meson production besides the normal quark combination at hadronization.
 Accordingly, the final distribution of $\phi$ meson is modeled
by incorporating both production mechanisms
\[
f_{\phi}^{(final)}(p_{T},\varphi)=f_{\phi,s\bar{s}}(p_{T},\varphi)+f_{\phi,KK}(p_{T},\varphi)\text{.}
\]
The elliptic flow is 
\begin{equation}
v_{2,\phi}^{(final)}(p_{T})=(1-z)v_{2,s\bar{s}}(p_{T})+zv_{2,KK}(p_{T}),\label{eq:phi_final}
\end{equation}
where the parameter $z$ denotes the fractional contribution of the
two-kaon coalescence channel and is given by $z=\frac{f_{\phi,KK}(p_{T})}{f_{\phi,s\bar{s}}(p_{T})+f_{\phi,KK}(p_{T})}$. 

\begin{figure}
\centering{}\includegraphics[width=0.85\linewidth,viewport=0bp 00bp 580bp 330bp]{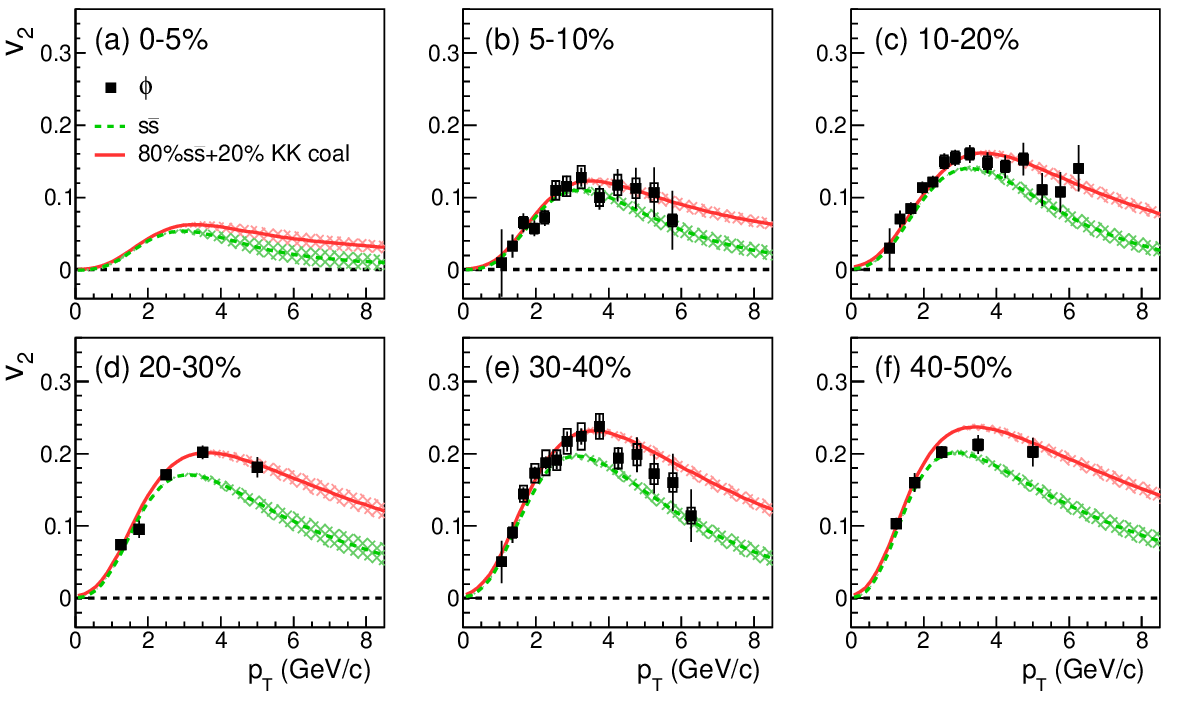}\caption{The $v_{2}$ of $\phi$ mesons as a function of $p_{T}$ in Pb+Pb
collisions at $\sqrt{s_{NN}}=$5.02 TeV. Experimental data (symbols)
are from Refs. \citep{ALICE:2022zks,Zhu:2019twz}. Theoretical results
including only the $s\overline{s}$ combination are shown as dashed
lines and those incorporating both $s\overline{s}$ combination and
two-kaon coalescence are shown as solid lines.\label{fig:fig4}}
\end{figure}

In Fig. \ref{fig:fig4}, we apply Eq. (\ref{eq:phi_final}) to study
$v_{2}$ of $\phi$ mesons and compare theoretical calculations with
experimental data \citep{ALICE:2022zks}. The results for the pure
$s\overline{s}$ quark combination mechanism, labeled as ``$s\overline{s}$'',
are shown as dashed curves in Fig. \ref{fig:fig4}. We see that this
mechanism can describe the $v_{2}$ of the $\phi$ mesons in the low
transverse momentum region ($p_{T}\lesssim2.5$ GeV/c) but underestimates
the data in the intermediate transverse momentum region ($p_{T}\apprge2.5$
GeV/c). If we introduce the channel of two-kaon coalescence ($KK\rightarrow\phi$)
with a contribution fraction $z=0.2$, we find that the final results
(the solid lines) are in good agreement with experimental data in
5-10\%, 10-20\%, and 20-30\% centralities in Pb+Pb collisions at $\sqrt{s_{NN}}=$5.02
TeV. This indicates that approximately 20\% of $\phi$ meson comes
from the two-kaon coalescence at hadronic rescattering state in Pb+Pb
collisions at $\sqrt{s_{NN}}=$5.02 TeV. In semi-central collisions
such as 30-40\% and 40-50\% centralities shown in Fig. \ref{fig:fig4}
(e) and (f), 20\% contribution from two-kaon coalescence is slightly
overestimated and 15\% is more fitted to experimental data. 

\subsection{THE TRIANGULAR FLOW OF LIGHT-FLAVOR HADRONS\label{subsec:light_hadron_v3}}

We further study the $v_{3}$ of light-flavor hadrons $\phi$, $\Lambda$,
$\Xi$, $\Omega,$ and $p$ at mid-rapidity in different centralities
in Pb+Pb collisions at $\sqrt{s_{NN}}=$5.02 TeV \citep{ALICE:2018yph,Zhu:2019twz}.
$v_{3}$ of light-flavor hadrons in Eqs. (\ref{eq:v3m_final}) and
(\ref{eq:v3b_final}) contains the contribution of the quadratic term
of quark $v_{2}$. According to the NCQ estimation of the $2-5^{th}$
flow of hadrons \citep{PHENIX:2006dpn,STAR:2015gge,ALICE:2014wao,ALICE:2018yph,STAR:2013ayu},
the $a_{5}/a_{3}$ is about 2.5 \citep{Kolb:2004gi,Chen:2004dv,STAR:2004jwm,STAR:2007afq,Bairathi:2016dob}.
Therefore, the magnitude of the $(a_{5}/a_{3}-2)v_{2,q}^{2}$ is about
$10^{-3}$ and its influence can be neglected in the formula, and
we finally obtain 
\begin{align}
v_{3,M_{j}}\left(p_{T}\right) & \approx v_{3,q_{1}}\left(x_{1}p_{T}\right)+v_{3,\bar{q}_{2}}\left(x_{2}p_{T}\right),\label{eq:v3,Mi_simp}\\
v_{3,B_{j}}\left(p_{T}\right) & \approx v_{3,q_{1}}\left(x_{1}p_{T}\right)+v_{3,q_{2}}\left(x_{2}p_{T}\right)+v_{3,q_{3}}\left(x_{3}p_{T}\right).\label{eq:v3,Bi_simp}
\end{align}
Applying Eqs. (\ref{eq:v3,Mi_simp}$-$\ref{eq:v3,Bi_simp}), we obtain
\begin{align}
v_{3,p}\left(p_{T}\right) & =3v_{3,u}\left(p_{T}/3\right),\label{eq:v3,p}\\
v_{3,\Omega}\left(p_{T}\right) & =3v_{3,s}\left(p_{T}/3\right),\label{eq:v3,omg}\\
v_{3,\phi}\left(p_{T}\right) & =v_{3,s}\left(p_{T}/2\right)+v_{3,\overline{s}}\left(p_{T}/2\right),\label{eq:v3,phi}\\
v_{3,\Lambda}\left(p_{T}\right) & =2v_{3,u}\bigl(\frac{1}{2+r_{su}}p_{T}\bigr)+v_{3,s}\bigl(\frac{r_{su}}{2+r_{su}}p_{T}\bigr),\label{eq:v3,lamb}\\
v_{3,\Xi}\left(p_{T}\right) & =2v_{3,s}\bigl(\frac{r_{su}}{1+2r_{su}}p_{T}\bigr)+v_{3,u}\bigl(\frac{1}{1+2r_{su}}p_{T}\bigr).\label{eq:v3,xi}
\end{align}
Here, we also assume the approximate isospin symmetry ($v_{3,u}=v_{3,d}$
) and strangeness neutrality $(v_{3,s}=v_{3,\bar{s}}$) at mid-rapidity
in Pb+Pb collisions at LHC energies.

\begin{figure}
\centering{}\includegraphics[width=0.7\linewidth,viewport=0bp 00bp 620bp 260bp]{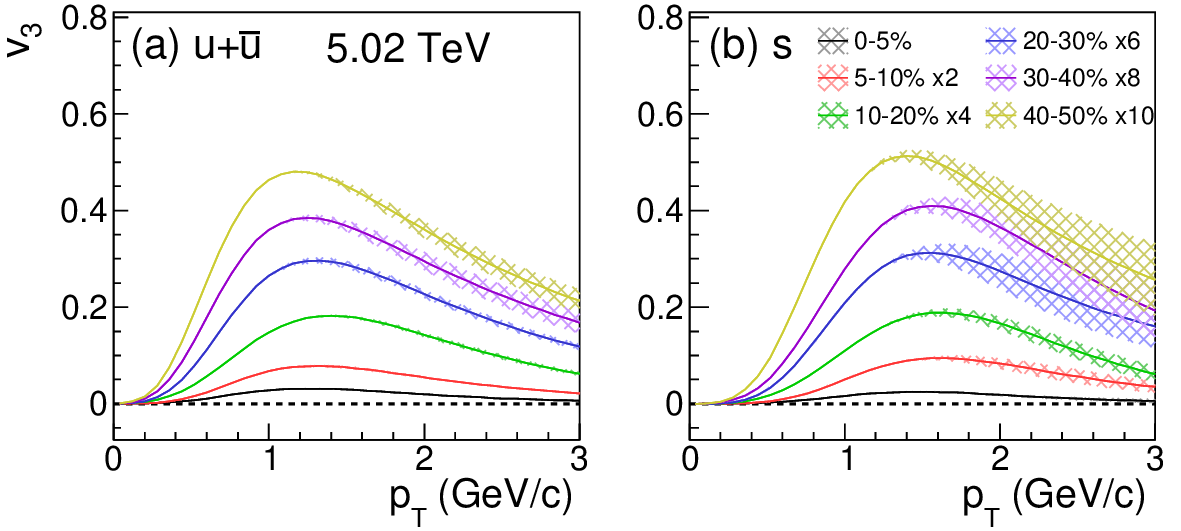}\caption{$v_{3}$ of quarks as a function of $p_{T}$ in different centralities
in Pb+Pb collisions at $\sqrt{s_{NN}}=$ 5.02 TeV.\textcolor{magenta}{{}
\label{fig:fig5}}}
\end{figure}

\begin{figure}
\centering{}\includegraphics[width=0.85\linewidth,viewport=0bp 00bp 580bp 350bp]{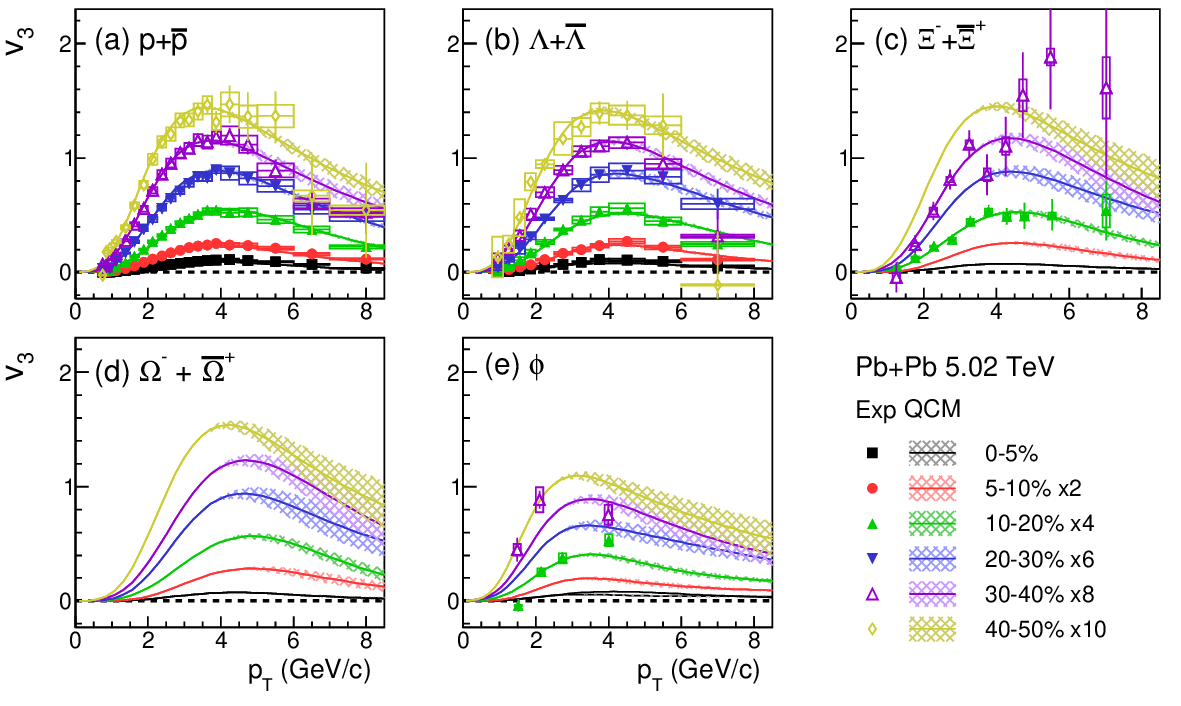}\caption{The $v_{3}$ of (a) $p$, (b) $\Lambda$, (c) $\Xi$, (d) $\Omega,$
and (e) $\phi$ in different centralities in Pb+Pb collisions at $\sqrt{s_{NN}}=$5.02
TeV. Symbols are the experimental data \citep{ALICE:2018yph,Zhu:2019twz}
and lines are model results.\label{fig:fig6}}
\end{figure}

We adopt an approach similar to that used for the hadronic $v_{2}$
in the previous section to test EVC model. First, we adopt the formula
in Eq. (\ref{eq:v2q}) to parameterize the $v_{3}$ of (anti-)quarks,
and then extract the up/down quark $v_{3,u}$ and strange quark $v_{3,s}$
by fitting experimental data of $p$ and $\Lambda$ measured in Pb+Pb
collisions at LHC energies. Results are shown in Fig. \ref{fig:fig5},
where data from different centralities are scaled by different factors
for clarity. We further use the obtained $v_{3,q}$ to calculate the
$v_{3}$ of other hadrons (such as $\Xi$, $\Omega,$ and $\phi$)
and compare them with corresponding experimental data. Fig. \ref{fig:fig6}
shows the results for the $v_{3}$ of $\phi$, $\Lambda$, $\Xi$,
$\Omega$, and $p$ in 0-5\%, 5-10\%, 10-20\%, 20-30\%, 30-40\%, 40-50\%
centralities in Pb+Pb collisions at $\sqrt{s_{NN}}=$ 5.02 TeV. Experimental
data of the $v_{3}$ of these hadrons are shown as symbols in Fig.
\ref{fig:fig6} and are taken from Refs. \citep{ALICE:2018yph,Zhu:2019twz}.
In the calculation of $\phi$ mesons, we also consider the contribution
of two-kaon coalescence to $v_{3}$ of $\phi$ and set $z=0.2$ in
the final prediction. 

Compared with the data points of $v_{2}$ for hadrons, the data points
of $v_{3}$ for hadrons are relatively fewer. By comparing between
the theoretical results and those limited experimental data, we see
that theoretical calculations of $v_{3}$ of $\Xi$ and $\phi$ are
generally in agreement with the available experimental data in the
studied collision centralities.

\section{THE ANISOTROPY FLOW OF SINGLE-CHARMED HADRONS\label{sec:charm hadron} }

In this section, we study the $v_{2}$ and $v_{3}$ of single-charmed
hadrons $D^{0}$, $D_{s}^{+}$, $\Lambda_{c}^{+}$, $\Xi_{c}^{0}$,
$\Omega_{c}^{0}$ at mid-rapidity in different centralities (0-10\%,
10-30\%, 30-50\%) in Pb+Pb collisions at $\sqrt{s_{NN}}=$2.76 and
5.02 TeV. 

\subsection{THE ELLIPTIC FLOW OF SINGLE-CHARMED HADRONS}

The $v_{2}$ of single-charmed hadrons is also given by Eqs. (\ref{eq:v2m_final})
and (\ref{eq:v2b_final}). For the light-flavor quarks, the scaling
coefficient $a_{4}$ is approximated as 2 according to our previous
analysis \citep{Feng:2025wde}. For charm quarks, although the exact
value of $a_{4}$ is not determined, the magnitude of the $(a_{4}-2)v_{2,q}^{2}$
is estimated to be less than 0.4\%. Therefore, we neglect the small
contribution of quadratic terms of quark $v_{2}$ and simplify the
$v_{2}$ of single-charmed hadrons as
\begin{align}
v_{2,M_{c\bar{l}}}\left(p_{T}\right) & \approx v_{2,q_{c}}\left(x_{1}p_{T}\right)+v_{2,q_{\bar{l}}}\left(x_{2}p_{T}\right),\label{eq:v2,Mi_charm}\\
v_{2,B_{cll^{\prime}}}\left(p_{T}\right) & \approx v_{2,q_{c}}\left(x_{1}p_{T}\right)+v_{2,q_{l}}\left(x_{2}p_{T}\right)+v_{2,q_{l^{\prime}}}\left(x_{3}p_{T}\right).\label{eq:v2,Bi_charm}
\end{align}
We observe a simple sum rule for the $v_{2}$ of charmed hadrons,
which is convenient for experimental testing. Under the assumption
of approximate isospin symmetry between up and down quarks (i.e.,
$v_{2,u}=v_{2,d}$), and strangeness neutrality (i.e., $v_{2,s}=v_{2,\bar{s}}$),
the formulas for $D^{0}$, $D_{s}^{+}$, $\Lambda_{c}^{+}$, $\Xi_{c}^{0},$
and $\Omega_{c}^{0}$ can be further simplified using Eqs. (\ref{eq:v2,Mi_charm})
and (\ref{eq:v2,Bi_charm}) as follows
\begin{align}
v_{2,D}\left(p_{T}\right) & =v_{2,c}\left(\frac{r_{cu}}{1+r_{cu}}p_{T}\right)+v_{2,u}\left(\frac{1}{1+r_{cu}}p_{T}\right),\label{eq:v2,D}\\
v_{2,D_{s}}\left(p_{T}\right) & =v_{2,c}\bigl(\frac{r_{cs}}{1+r_{cs}}p_{T}\bigr)+v_{2,s}\bigl(\frac{1}{1+r_{cs}}p_{T}\bigr),\label{eq:v2,Ds}\\
v_{2,\Lambda_{c}}\left(p_{T}\right) & =v_{2,c}\bigl(\frac{r_{cu}}{2+r_{cu}}p_{T}\bigr)+2v_{2,u}\bigl(\frac{1}{2+r_{cu}}p_{T}\bigr),\label{eq:v2,lamb_c}\\
v_{2,\Xi_{c}}\left(p_{T}\right) & =v_{2,c}\bigl(\frac{r_{cu}}{1+r_{su}+r_{cu}}p_{T}\bigr)+v_{2,u}\bigl(\frac{1}{1+r_{su}+r_{cu}}p_{T}\bigr)+v_{2,s}\bigl(\frac{r_{su}}{1+r_{su}+r_{cu}}p_{T}\bigr),\label{eq:v2,Xi_c}\\
v_{2,\Omega_{c}}\left(p_{T}\right) & =v_{2,c}\bigl(\frac{r_{cs}}{2+r_{cs}}p_{T}\bigr)+2v_{2,s}\bigl(\frac{1}{2+r_{cs}}p_{T}\bigr).\label{eq:v2,Omg_c}
\end{align}
Here, the relative momentum ratios are $r_{cu}=m_{c}/m_{u}=5$ and
$r_{cs}=m_{c}/m_{s}=3.$ It can be seen that the $v_{2}$ of charmed
hadrons at $p_{T}$ is the summation of the $v_{2}$ of the charm
quark whose $p_{T}$ is close to that of the charmed hadron and the
$v_{2}$ of a light-flavor quark whose $p_{T}$ is much smaller. This
composition pattern is different from that of light-flavor hadrons
where quarks have relatively close momentum. 

\begin{figure}
\centering{}\includegraphics[width=0.7\linewidth,viewport=0bp 00bp 620bp 260bp]{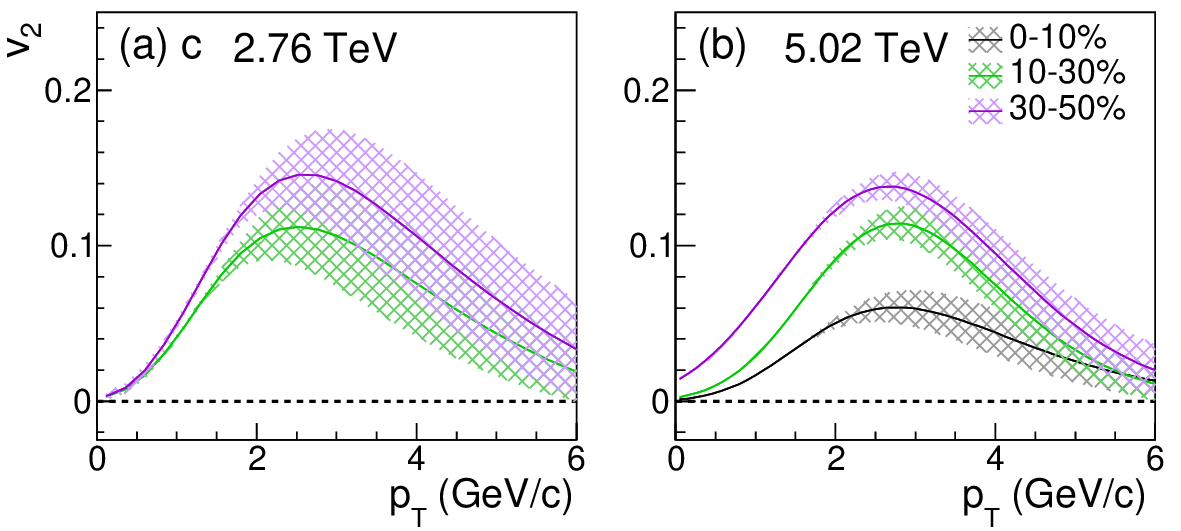}\caption{$v_{2}$ of $c$ quarks as a function of $p_{T}$ in different centralities
in Pb+Pb collisions at $\sqrt{s_{NN}}=$2.76 and 5.02 TeV. \label{fig:fig7}}
\end{figure}

Given the $v_{2,u}$ and $v_{2,s}$ obtained in the previous Sec.~
\ref{sec:light hadron_v2}, the $v_{2}$ of charm quarks can be extracted
from the experimental data of $D$ mesons, 
\begin{equation}
v_{2,c}(p_{T})=v_{2,D}\left(\frac{1+r_{cu}}{r_{cu}}p_{T}\right)-v_{2,u}\left(\frac{1}{r_{cu}}p_{T}\right).\label{eq:v2,c}
\end{equation}
For the charm quark, the parameterization form also adopts Eq. (\ref{eq:v2q}).
In Fig. \ref{fig:fig7}, we show the extracted $v_{2,c}$ as a function
of $p_{T}$ in different centralities in Pb+Pb collisions at $\sqrt{s_{NN}}=$2.76
and 5.02 TeV. The shadow areas show the statistical uncertainties,
which are passed from the statistical uncertainties of the experimental
data for $D^{0}$ \citep{Bailhache:2014fia,CMS:2017vhp,ALICE:2020pvw}.
It should be noted that this extraction method is only valid in the
low and intermediate $p_{T}$ region, where the combination mechanism
dominates $D$ meson production. In previous studies \citep{Song:2023nzu,Chang:2023zbe,Li:2017zuj,Song:2018tpv},
we found that the experimental data for $p_{T}$ spectra of single-charmed
hadrons in the $p_{T}\apprle$ 7.5 GeV/c range from pp, p+Pb, and
Pb+Pb collisions at LHC energies are well described by the EVC model.
Therefore, the experimental data of $v_{2}$ for $D$ mesons with
$p_{T}\apprle$ 7.5 GeV/c can be used to extract the $v_{2}$ of charm
quarks with $p_{T}\apprle$ 6 GeV/c within the EVC model. In the higher
$p_{T}$ region ($p_{T}\apprge$ 7.5 GeV/c), the fragmentation mechanism
becomes significant \citep{Oh:2009zj,Cao:2015hia}, and Eq. (\ref{eq:v2,c})
is no longer applicable.

\begin{figure}
\centering{}\includegraphics[width=0.85\linewidth,viewport=0bp 00bp 580bp 350bp]{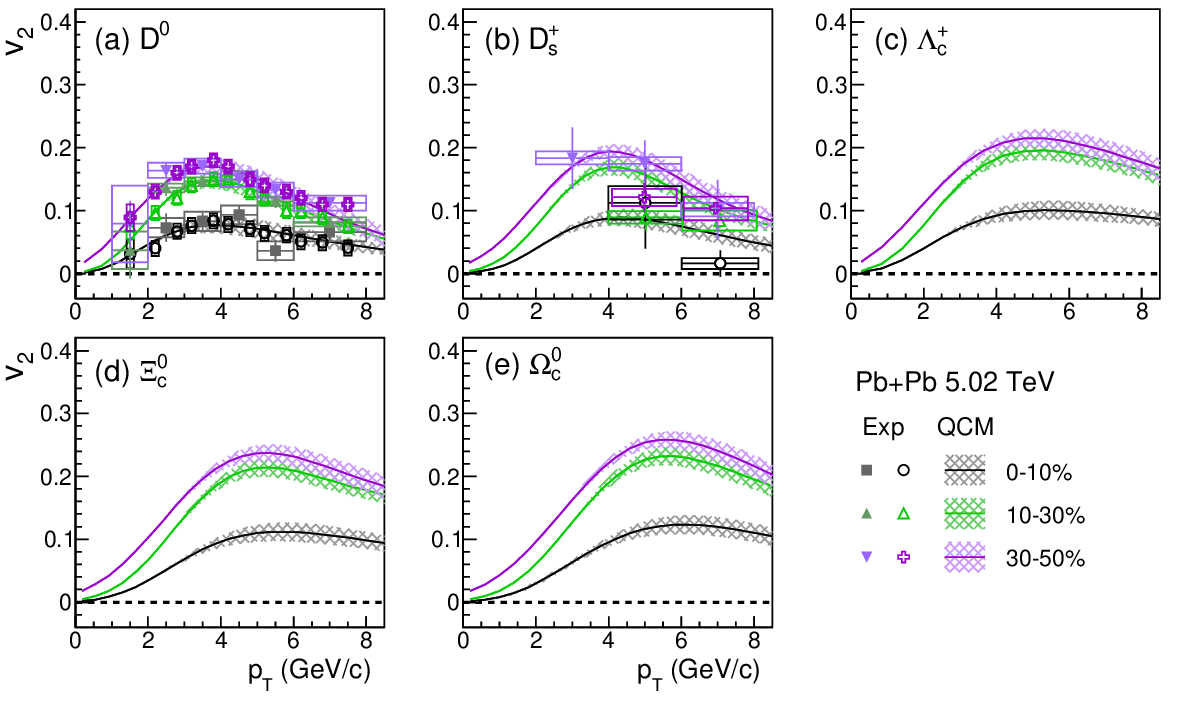}\caption{The $v_{2}$ of (a) $D^{0}$, (b) $D_{s}^{+}$, (c) $\Lambda_{c}^{+}$
, (d) $\Xi_{c}^{0}$, and (e) $\Omega_{c}^{0}$ in different centralities
in Pb+Pb collisions at $\sqrt{s_{NN}}=$5.02 TeV. Symbols are the
experimental data \citep{CMS:2020bnz,ALICE:2020pvw,ALICE:2021kfc,CMS:2025cdf}
and lines are model results. \label{fig:fig8}}
\end{figure}

\begin{figure}
\centering{}\includegraphics[width=0.85\linewidth,viewport=0bp 00bp 580bp 350bp]{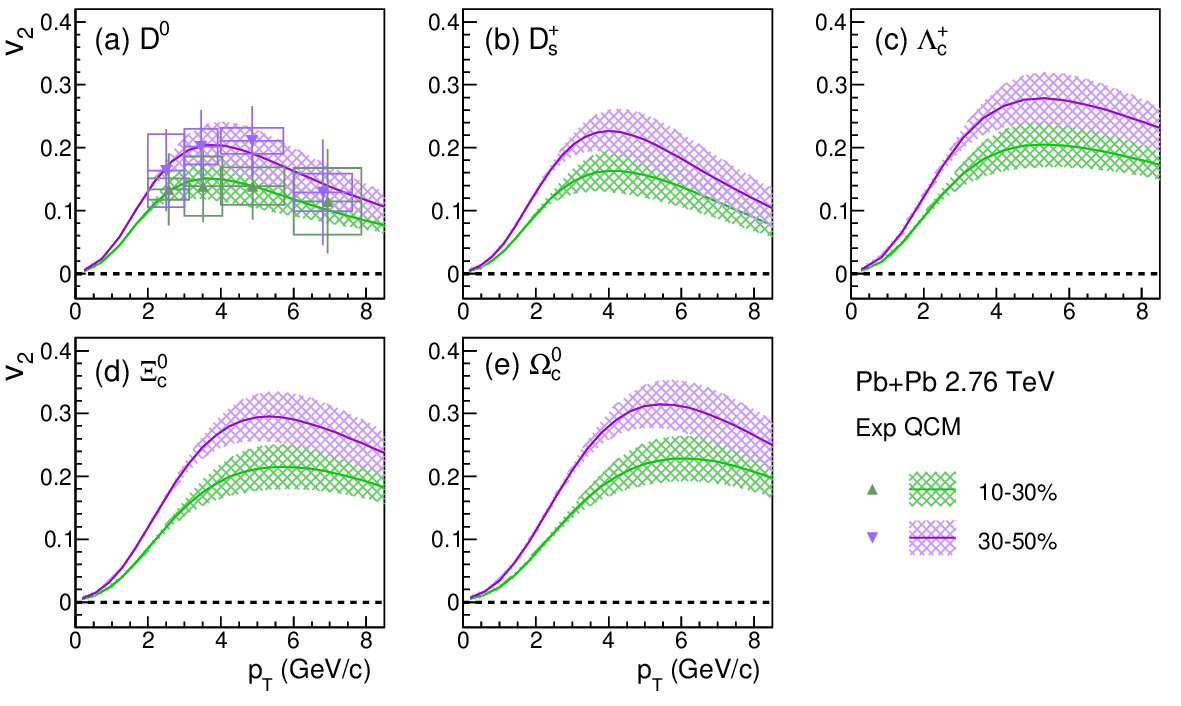}\caption{The $v_{2}$ of (a) $D^{0}$, (b) $D_{s}^{+}$, (c) $\Lambda_{c}^{+}$
, (d) $\Xi_{c}^{0},$ and (e) $\Omega_{c}^{0}$ in different centralities
in Pb+Pb collisions at $\sqrt{s_{NN}}=$2.76 TeV. Symbols are the
experimental data \citep{Bailhache:2014fia} and lines are model results.
\label{fig:fig9}}
\end{figure}

\begin{figure}
\centering{}\includegraphics[width=0.65\linewidth,viewport=0bp 00bp 580bp 450bp]{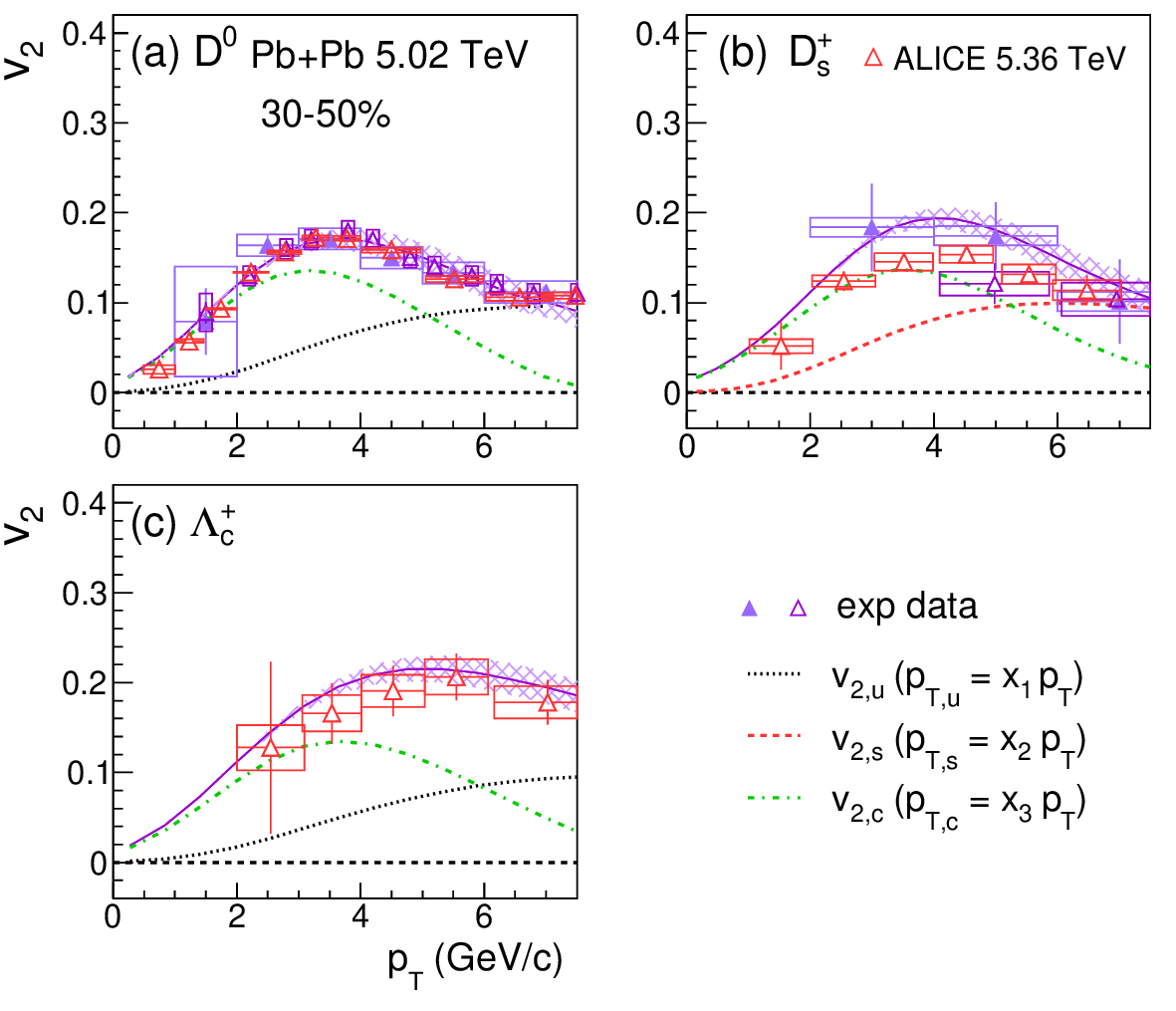}\caption{The $v_{2}$ of (a) $D^{0}$, (b) $D_{s}^{+}$, and (c) $\Lambda_{c}^{+}$
in 30-50\% centrality in Pb+Pb collisions at $\sqrt{s_{NN}}=$5.02
and 5.36 TeV. Symbols are the experimental data \citep{CMS:2020bnz,ALICE:2020pvw,ALICE:2021kfc,CMS:2025cdf}
and lines are model results. \label{fig:fig10}}
\end{figure}

In Figs. \ref{fig:fig8}-\ref{fig:fig9}, we present the results for
the $v_{2}$ of $D^{0}$, $D_{s}^{+}$, $\Lambda_{c}^{+}$, $\Xi_{c}^{0}$,
and $\Omega_{c}^{0}$ at 0-10\%, 10-30\%, and 30-50\% centralities
in Pb+Pb collisions at $\sqrt{s_{NN}}=$2.76 and 5.02 TeV. The symbols
are the experimental data \citep{CMS:2020bnz,ALICE:2020pvw,ALICE:2021kfc,CMS:2025cdf,Bailhache:2014fia}
from the CMS and ALICE collaborations and lines are the calculated
results. The results are in good agreement with the experimental data
for $D_{s}^{+}$ in Fig. \ref{fig:fig8}(b). Predictions from QCM
for other charmed baryons $\Lambda_{c}^{+}$, $\Xi_{c}^{0}$ , $\Omega_{c}^{0}$
are presented in Figs. \ref{fig:fig8}(b)-(d) and \ref{fig:fig9}(b)-(d),
which can be tested in future experimental measurements. 

Recently, ALICE collaboration report the preliminary data for $v_{2}$
of single-charmed hadrons $D^{0}$, $D_{s}^{+},$ and $\Lambda_{c}^{+}$
in Pb+Pb collisions at $\sqrt{s_{NN}}=$ 5.36 TeV. Because of quite
close collision energy, these data can be used as the first test of
our prediction. In Fig. \ref{fig:fig10} we show these data in 30-50\%
centrality as well as $v_{2}$ of $D^{0}$ and $D_{s}^{+}$ at $\sqrt{s_{NN}}=$
5.02 TeV. We see that $v_{2}$ of $D^{0}$ at the two collision energies
are quite close with each other and $v_{2}$ of ${\color{magenta}D_{s}^{+}}$
has a slight difference but the error bar at 5.02 TeV is large. 

In order to clearly show the individual contributions of charm quarks
and light-flavor quarks in the composition of $v_{2}$ of charmed
hadrons, we plot the contribution of charm quarks as the dot-dashed
lines and that of light-flavor quarks as the dotted lines. The solid
lines represent $v_{2}$ of charmed hadrons in Fig. \ref{fig:fig10}
(a)-(c). From Fig. \ref{fig:fig10} (a), we see that $v_{2}$ of $D^{0}$
in the range $p_{T}\lesssim2$ GeV/c is dominated by that of charm
quarks and at large $p_{T}$ the contribution of light-flavor quarks
continuously strengths and is even dominated at $p_{T}\gtrsim5$ GeV/c.
$v_{2}$ of $D_{s}^{+}$ and $\Lambda_{c}^{+}$ in Fig. \ref{fig:fig10}
(b)-(c) also exhibit similar property. A difference between $\Lambda_{c}^{+}$
and mesons $D_{s}^{+}$ or $D^{0}$ is that $\Lambda_{c}^{+}$ contains
double contribution from light-flavor quark $v_{2}$. Therefore, the
$v_{2}$ of $\Lambda_{c}^{+}$ is larger than those of $D_{s}^{+}$
or $D^{0}$ at intermediate $p_{T}$ but it does not follow the NCQ
scaling exhibited in light-flavor hadrons because $v_{2}$ of involved
light-flavor quarks is quite different from $v_{2}$ of charm quarks,
see the crossing feature between the dotted line and dot-dashed line
in Fig. \ref{fig:fig10} (a)-(c). Another related property for $v_{2}$
of charmed meson and charmed baryon is that the peak position for
$v_{2}$ of $\Lambda_{c}^{+}$ is at about 5 GeV/c while that of $D^{0}$
is at about $3.5$ GeV. This is because the double contribution from
light-flavor quarks in $\Lambda_{c}^{+}$ formation will push the
peak position for $v_{2}$ of $\Lambda_{c}^{+}$ to larger $p_{T}$
to a certain extent.

\subsection{THE TRIANGULAR FLOW OF SINGLE-CHARMED HADRONS}

In this subsection, we study the $v_{3}$ of single-charmed hadrons
$D^{0}$, $D_{s}^{+}$, $\Lambda_{c}^{+}$, $\Xi_{c}^{0}$, and $\Omega_{c}^{0}$
in different centralities in Pb+Pb collisions at $\sqrt{s_{NN}}=$
5.02 TeV, and make comparisons with the available experimental data
\citep{CMS:2025cdf,CMS:2017vhp,ALICE:2020iug}. $v_{3}$ of single-charmed
hadrons in Eqs. (\ref{eq:v3m_final}) and (\ref{eq:v3b_final}) includes
the small contribution from the quadratic terms of quark $v_{2}$.
By NCQ scaling estimation for the second to fifth harmonic flows of
light-flavor hadrons \citep{PHENIX:2006dpn,STAR:2015gge,ALICE:2014wao,ALICE:2018yph,STAR:2013ayu},
the relating coefficient $\frac{a_{5}}{a_{3}}$ is approximately 2.5
\citep{Kolb:2004gi,Chen:2004dv,STAR:2004jwm,STAR:2007afq,Bairathi:2016dob}.
Consequently, the magnitude of $\left(\frac{a_{5}}{a_{3}}-2\right)v_{2,q}^{2}$
is on the order of $10^{-3}$, and its influence is negligible in
the formula. We therefore finally obtain
\begin{align}
v_{3,M_{c\bar{l}}}\left(p_{T}\right) & \approx v_{3,q_{c}}\left(x_{1}p_{T}\right)+v_{3,q_{\bar{l}}}\left(x_{2}p_{T}\right),\label{eq:v3,Mi_charm}\\
v_{3,B_{cll^{\prime}}}\left(p_{T}\right) & \approx v_{3,q_{c}}\left(x_{1}p_{T}\right)+v_{3,q_{l}}\left(x_{2}p_{T}\right)+v_{3,q_{l^{\prime}}}\left(x_{3}p_{T}\right),\label{eq:v3,Bi_charm}
\end{align}
and obtain 
\begin{align}
v_{3,D}\left(p_{T}\right) & =v_{3,c}\left(\frac{r_{cu}}{1+r_{cu}}p_{T}\right)+v_{3,u}\left(\frac{1}{1+r_{cu}}p_{T}\right),\label{eq:v3,D}\\
v_{3,D_{s}}\left(p_{T}\right) & =v_{3,c}\bigl(\frac{r_{cs}}{1+r_{cs}}p_{T}\bigr)+v_{3,s}\bigl(\frac{1}{1+r_{cs}}p_{T}\bigr),\label{eq:v3,Ds}\\
v_{3,\Lambda_{c}}\left(p_{T}\right) & =v_{3,c}\bigl(\frac{r_{cu}}{2+r_{cu}}p_{T}\bigr)+2v_{3,u}\bigl(\frac{1}{2+r_{cu}}p_{T}\bigr),\label{eq:v3,lamb_c}\\
v_{3,\Xi_{c}}\left(p_{T}\right) & =v_{3,c}\bigl(\frac{r_{cu}}{1+r_{su}+r_{cu}}p_{T}\bigr)+v_{3,u}\bigl(\frac{1}{1+r_{su}+r_{cu}}p_{T}\bigr)+v_{3,s}\bigl(\frac{r_{su}}{1+r_{su}+r_{cu}}p_{T}\bigr),\label{eq:v3,Xi_c}\\
v_{3,\Omega_{c}}\left(p_{T}\right) & =v_{3,c}\bigl(\frac{r_{cs}}{2+r_{cs}}p_{T}\bigr)+2v_{3,s}\bigl(\frac{1}{2+r_{cs}}p_{T}\bigr).\label{eq:v3,Omg_c}
\end{align}
Here, we also adopt approximate isospin symmetry between up and down
quarks, as well as strangeness neutrality between strange quarks and
strange antiquarks at mid-rapidity in heavy-ion collisions at LHC
energies.

\begin{figure}
\centering{}\includegraphics[width=0.45\linewidth,viewport=0bp 00bp 620bp 510bp]{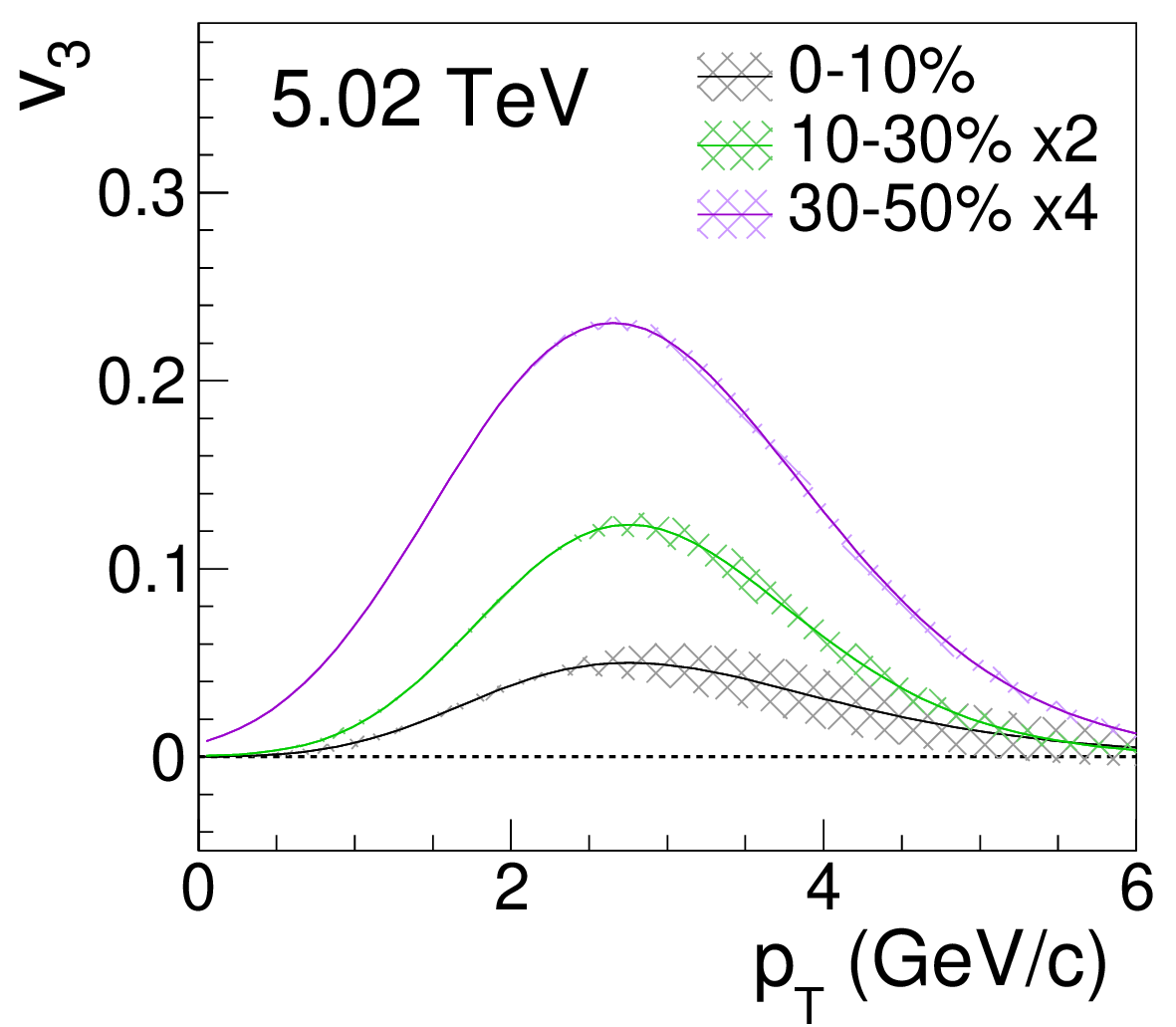}\caption{$v_{3}$ of $c$ quarks as a function of $p_{T}$ in different centralities
in Pb+Pb collisions at $\sqrt{s_{NN}}=$ 5.02 TeV. \label{fig:fig11}}
\end{figure}

We employ the procedure similar to that used in testing $v_{2}$ of
the single-charmed hadrons in the previous section to examine the
$v_{3}$ of the single-charmed hadrons. First, we again adopt Eq.
(\ref{eq:v2q}) to parameterize the $v_{3}$ of charm quarks. By fitting
experimental data from $D^{0}$ in Pb+Pb collisions at $\sqrt{s_{NN}}=$
5.02 TeV, the values of the parameters $v_{3,c}$ are extracted. Figure
\ref{fig:fig11} shows the extracted $v_{3,c}$ in the three centralities,
where data from different centralities are scaled by different factors
for clarity. Second, we use the resulting $v_{3,c}$ and $v_{3}$
of light-flavor quarks (i.e., $v_{3,u},$$v_{3,s}$) obtained in Sec.
\ref{subsec:light_hadron_v3} to calculate the $v_{3}$ of other single-charmed
hadrons, such as $D_{s}^{+}$, $\Lambda_{c}^{+}$, $\Xi_{c}^{0}$,
and $\Omega_{c}^{0}$ by Eqs. (\ref{eq:v3,Ds})-(\ref{eq:v3,Omg_c}),
and compare them with experimental data.

\begin{figure}
\centering{}\includegraphics[width=0.85\linewidth,viewport=0bp 00bp 580bp 350bp]{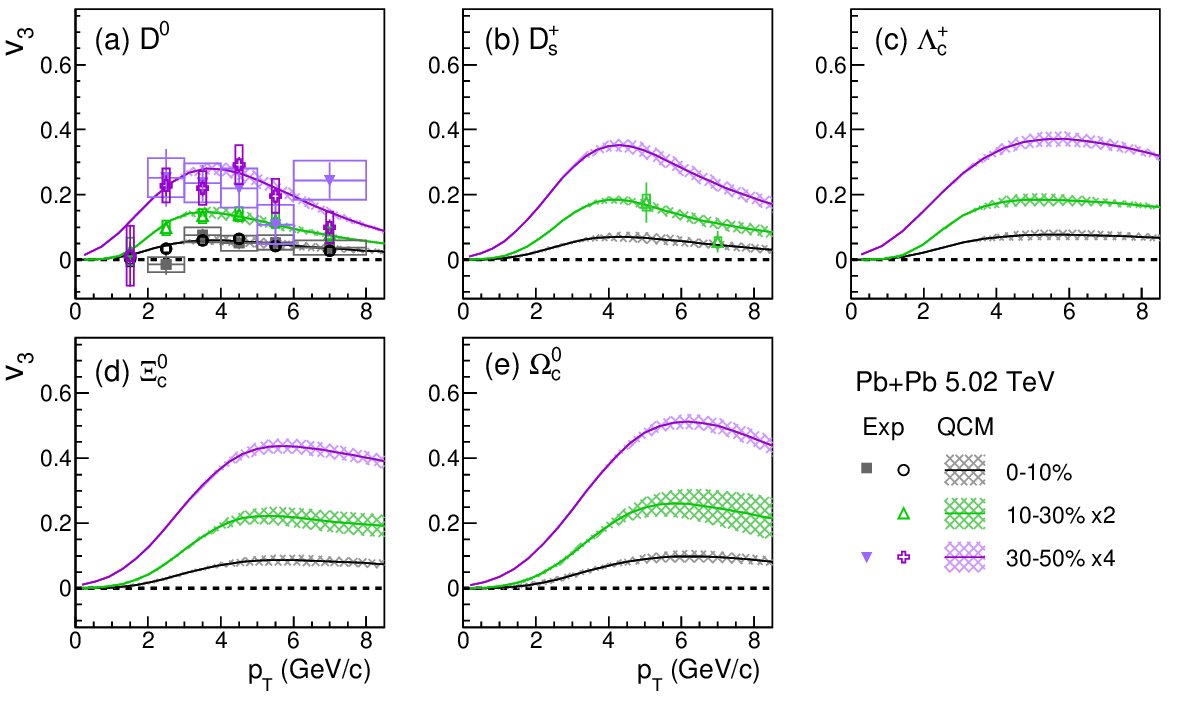}\caption{The $v_{3}$ of (a) $D^{0}$, (b) $D_{s}^{+}$, (c) $\Lambda_{c}^{+}$,
(d) $\Xi_{c}^{0}$, and (e) $\Omega_{c}^{0}$ in different centralities
in Pb+Pb collisions at $\sqrt{s_{NN}}=$5.02 TeV. Symbols are the
experimental data \citep{CMS:2025cdf,CMS:2017vhp,ALICE:2020iug} and
lines are model results. \label{fig:fig12}}
\end{figure}

Fig. \ref{fig:fig12} shows the results of the $v_{3}$ of $D^{0}$,
$D_{s}^{+}$, $\Lambda_{c}^{+}$, $\Xi_{c}^{0}$, and $\Omega_{c}^{0}$
in Pb+Pb collisions at $\sqrt{s_{NN}}=$5.02 TeV in the 0-10\%, 10-30\%,
and 30-50\% centralities. Comparing Eqs. (\ref{eq:v3,D})-(\ref{eq:v3,Omg_c})
for $v_{3}$ and Eqs. (\ref{eq:v2,D})-(\ref{eq:v2,Omg_c}) for $v_{2}$
of charmed hadrons, we see that the composition pattern of charm quarks
and light-flavor quarks are very similar in $v_{2}$ and $v_{3}$
of charmed hadrons. Therefore, like those shown in Fig. \ref{fig:fig10}
for hadronic $v_{2}$, $v_{3}$ of charmed hadrons in the range $p_{T}\lesssim4$
GeV/c is dominated by $v_{3}$ of charm quarks and in the intermediate
$p_{T}$ region ($4\lesssim p_{T}\lesssim7.5$ GeV/c), the contribution
from light-flavor quarks gradually increases. Taking results in 30-50\%
centrality as an example, we see that the peak position for $v_{3}$
of $D^{0}$, $D_{s}^{+}$ is located at about 4.0 GeV/c while that
of baryons $\Lambda_{c}^{+}$, $\Xi_{c}^{0}$, and $\Omega_{c}^{0}$
is located at about 5.5 GeV/c. The maximum $v_{3}$ value of $\Lambda_{c}^{+}$
at peak position is higher than that of $D^{0}$ about 0.07 because
of an extra contribution of light-flavor quark $v_{3}$. The maximum
$v_{3}$ of strange charmed hadrons $D_{s}^{+}$ , $\Xi_{c}^{0}$,
and $\Omega_{c}^{0}$ are higher than those of $D^{0}$ and $\Lambda_{c}^{+}$
because of the involved strange quarks have larger transverse momentum
and thus larger $v_{3}$ at charm quark hadronization. These properties
of $v_{3}$ of charmed hadrons are left for future test.

\subsection{THE PROPERTIES OF $v_{2}$ and $v_{3}$ OF QUARKS}

In this subsection, we utilize the previously obtained $v_{2}$ and
$v_{3}$ of up/down quarks, strange quarks and charm quarks just before
hadronization to further investigate their properties, with a particular
focus on the scaling behavior of anisotropic flow at different orders.

\begin{figure}
\begin{centering}
\includegraphics[width=0.65\linewidth,viewport=0bp 00bp 580bp 500bp]{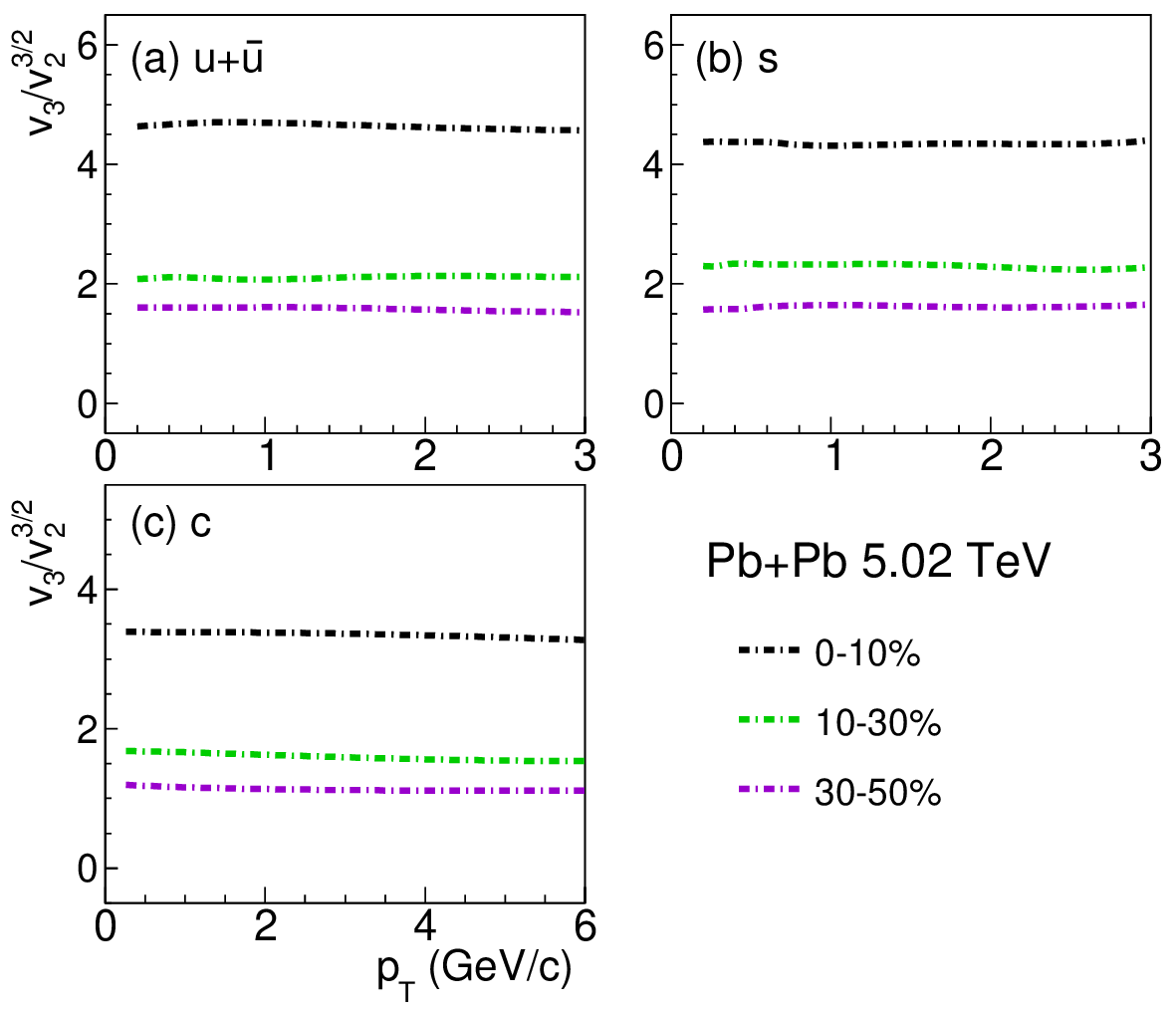}
\par\end{centering}
\caption{The ratio $v_{3}/v_{2}^{3/2}$ of quarks as a function of $p_{T}$
in different centralities in Pb+Pb collisions at $\sqrt{s_{NN}}=$5.02
TeV. \label{fig:fig13}}
\end{figure}

In Section \ref{sec:EVC_model}, we assumed a scaling relation $v_{n,q}=a_{n}v_{2,q}^{n/2}$
at quark level to estimate the contribution of higher-order quark
flows to the lower-order anisotropic flow of single-charmed hadrons.
Here, we test this relation using the obtained $v_{2}$ and $v_{3}$
of up/down, strange, and charm quarks. The results of $a_{3}=v_{3}/v_{2}^{3/2}$
as a function of $p_{T}$ are shown in Fig. \ref{fig:fig13}. We observe
that the values of $a_{3}$ for up, strange, and charm quarks all
exhibit a weak dependence on $p_{T}$, clustering around a value of
approximately 2$\sim4$. The same behavior is also seen at RHIC. Moreover,
the value of $a_{3}$ for charm quarks is observed to be significantly
smaller than those for light-flavor quarks in the three centralities.
This may be explained by the observation that heavier quarks exhibit
a more pronounced centrality dependence in their elliptic flow, whereas
triangular flow shows relatively little sensitivity to centrality
variations. These results validate the consistency of the scaling
relation $v_{n,q}=a_{n}v_{2,q}^{n/2}$ as used in the current model
for analyzing the anisotropic flow $v_{n}$ of single-charmed hadrons.
The observed scaling behavior of quark flows can be qualitatively
understood within the framework of hydrodynamic evolution of hot quark
matter \citep{Borghini:2005kd,Retinskaya:2013gca}.

\section{summary\label{sec:Summary}}

In this paper, we have applied the constituent quark equal-velocity
combination model to study $v_{2}$ and $v_{3}$ of light-flavor hadrons
and single-charmed hadrons in different centralities (0-10\%, 10-30\%,
and 30-50\%) in Pb+Pb collisions at $\sqrt{s_{NN}}=$2.76 and 5.02
TeV. In equal velocity combination mechanism, the $v_{2,3}$ of hadrons
can be finally expressed as the linear superposition of $v_{2,3}$
of constituent quarks and antiquarks after neglecting the higher order
flows of quarks. These simplified analytical expressions for hadronic
flows are conveniently tested by experimental data. 

Using these analytical expressions for hadronic flows, we firstly
systematically study the $v_{2}$ and $v_{3}$ of $p$, $\Lambda$,
$\Xi$, $\Omega$, and $\phi$ in different centralities in Pb+Pb
collisions at $\sqrt{s_{NN}}=$2.76 and 5.02 TeV. $v_{2}$ and $v_{3}$
of quarks are obtained by combination fitting to experimental data
of $\Lambda$ and $\Xi$, and we used them to calculate flows of other
hadrons and compare them with the available experimental data. We
found good agreement with data at $\sqrt{s_{NN}}=$2.76. In Pb+Pb
collisions at $\sqrt{s_{NN}}=$5.02 TeV, we found that $v_{2}$ data
of $\phi$ are higher than our predictions by $s\bar{s}$ quark combination.
We therefore considered the possible contribution of final-state hadronic
rescattering by including the production of $\phi$ via two-kaon coalescence.
We found that two-kaon coalescence channel contributes about 20\%
of $\phi$ production at $\sqrt{s_{NN}}=$5.02 TeV. With this contribution
estimation, we further predict $v_{3}$ of $\phi$ as the function
of $p_{T}$ for future test. 

Using the obtained $v_{2}$ and $v_{3}$ of light-flavor quarks in
studying light-flavor hadrons, we further studied the $v_{2}$ and
$v_{3}$ of single-charmed hadrons in Pb+Pb collisions at $\sqrt{s_{NN}}=$2.76
and 5.02 TeV. We used the experimental data of $D^{0}$ to fix the
$v_{2}$ and $v_{3}$ of charm quarks and then applied EVC model to
predicted flows of $D_{s}^{+}$, $\Lambda_{c}^{+}$, $\Xi_{c}^{0}$,
and $\Omega_{c}^{+}$ at two collision energies. We also took the
preliminary data of $D^{0}$, $D_{s}^{+}$, and $\Lambda_{c}^{+}$
in 30-50\% centrality in Pb+Pb collisions at $\sqrt{s_{NN}}=$5.36
TeV to compare with our predictions to show some features of $v_{2}$
of charmed hadron, see Fig. 10. An interesting feature is that $v_{2}$
of $\Lambda_{c}^{+}$ is only slightly higher than those of $D^{0}$
and $D_{s}^{+}$ in small and intermediate $p_{T}$ ranges, which
is different from the NCQ scaling property for light-flavor hadrons.
This is because in EVC model charm quark hadronizes by capturing a
light-flavor antiquark or two light-flavor quarks with much smaller
momentum $p_{T,q}=m_{q}/m_{c}p_{T,c}\approx0.2-0.3p_{T,c}$. $v_{2}$
of charmed hadrons at $p_{T}$ is therefore the summation of $v_{2}$
of charm quark at a slightly smaller $p_{T}$ and that of light-flavor
quarks at a much smaller $p_{T}$ which has quite small value at low
$p_{T}$. Another property is that the $p_{T}$ position for the maximum
$v_{2}$ of $\Lambda_{c}^{+}$ is larger than that of D mesons about
1 GeV/c. This is because in $\Lambda_{c}^{+}$ formation a charm quark
will capture two light-flavor quarks and $v_{2}$ of light-flavor
quarks involving $\Lambda_{c}^{+}$ formation is ascending function
with $p_{T}$. 

In deriving $v_{2}$ and $v_{3}$ of hadrons, we assumed a scaling
relation $v_{n,q}=a_{n}(v_{2,q})^{n/2}$ for quark flows. Using the
obtained $v_{2}$ and $v_{3}$ of light-flavor quarks and charm quarks,
we found that this relation holds well for $v_{3}$ of light-flavor
quarks and charm quarks at the two LHC energies. This was consistent
with our previous results at RHIC energies \citep{Feng:2025wde}.
The origin of this scaling relation for quark flows is related to
hydrodynamic expansion evolution of QGP created in relativistic heavy-ion
collisions.

\section{Acknowledgments}

This work was supported in part by Shandong Provincial Natural Science
Foundation (Grant Nos. ZR2025QC38 and ZR2025MS01), and the National
Natural Science Foundation of China under Grants No. 12375074 and
12175115.

\bibliographystyle{apsrev4-2}
\bibliography{ref}

\end{document}